\documentclass[a4paper,11pt]{article}
\pdfoutput=1
\usepackage{jcappub}
\usepackage[T1]{fontenc} 
\usepackage{hyperref}
\usepackage{graphics}
\usepackage{amssymb, amsmath}
\usepackage{bm}
\usepackage{xcolor,color,colortbl}
\usepackage{multirow,bigdelim}
\usepackage{bbold}
\usepackage{verbatim}

\author[a,c,d,e]{Alex Krolewski}
\author[b,a]{Simone Ferraro}
\author[a,b,c]{Martin White}

\affiliation[a]{Department of Physics, University of California, Berkeley, CA 94720}
\affiliation[b]{Physics Division, Lawrence Berkeley National Laboratory, Berkeley, CA}
\affiliation[c]{Department of Astronomy, University of California, Berkeley, CA 94720}
\affiliation[d]{AMTD Fellow, Waterloo Centre for Astrophysics, University of Waterloo, Waterloo ON N2L 3G1, Canada}
\affiliation[e]{Perimeter Institute for Theoretical Physics, 31 Caroline St. North, Waterloo, ON NL2 2Y5, Canada}

\emailAdd{akrolewski@perimeterinstitute.ca}
\emailAdd{sferraro@lbl.gov}
\emailAdd{mwhite@berkeley.edu}

\title{Cosmological constraints from unWISE and Planck CMB lensing tomography}

\keywords{cosmological parameters from LSS -- power spectrum -- CMB --
galaxy clustering}

\abstract{
A number of recent, low-redshift, lensing measurements hint at a universe in which the amplitude of lensing is lower than that predicted from the $\Lambda$CDM model fit to the data of the Planck CMB mission.
Here we use the auto- and cross-correlation signal of unWISE galaxies and Planck CMB lensing maps to infer cosmological parameters at low redshift.
In particular, we consider three unWISE samples (denoted as ``blue'', ``green'' and ``red'') at median redshifts $z \sim 0.6$, $1.1$ and 1.5, which fully cover the Dark Energy dominated era. Our cross-correlation measurements, with combined significance $S/N \sim 80$, are used to infer the amplitude of low-redshift fluctuations, $\sigma_8$; the fraction of matter in the Universe, $\Omega_m$; and the combination $S_8 \equiv \sigma_8 (\Omega_m / 0.3)^{0.5}$ to which these low-redshift lensing measurements are most sensitive.  The combination of blue, green and red samples gives a value $S_8=0.784\pm 0.015$, that is fully consistent with other low-redshift lensing measurements and in 2.4$\sigma$ tension with the CMB predictions from Planck. This is noteworthy, because CMB lensing probes the same physics as previous galaxy lensing measurements, but with very different systematics, thus providing an excellent complement to previous measurements.}

\begin{document}
\maketitle
\flushbottom

\section{Introduction}

Weak lensing of the CMB (see \cite{Lewis06,Hanson10} for reviews) is rapidly becoming one of our most powerful cosmological tools. This is because of its rapidly-increasing statistical power and the robustness that arises from the well characterized statistical properties and redshift of the source (i.e.\ the primary CMB). 
CMB lensing by itself is sensitive to a wide range of redshifts. When cross-correlating CMB lensing with low-redshift tracers (CMB lensing tomography) such as galaxies, we extract the information over the redshift range of interest (see for example \cite{Sherwin2012, Bleem2012, PlanckLens13} for some of the early work and \cite{PlanckLens18,Omori2018b, Marques:2019aug,Darwish21,Hang21,Kitanidis21,Alonso21,Yan21} for more recent analyses). Moreover, because of the different dependence on galaxy bias of the auto and cross-correlations, we can efficiently break the degeneracy between bias and the amplitude of fluctuations at low redshift, providing tight cosmological constraints on the low-redshift Universe.

In a companion paper \cite{Krolewski20}, we have presented the cross-correlation signal between the Planck 2018 CMB lensing maps \cite{PlanckLens18} and the infrared-detected unWISE galaxies \cite{Schlafly19}. The unWISE catalog is further split in three redshift bins at median redshift $z \sim 0.6$, $1.1$ and 1.5, which we call the ``blue'', ``green'' and ``red'' samples respectively, and a number of quality cuts are performed to maximize the uniformity in properties and the masking of spurious sources. In ref.\ \cite{Krolewski20}, we also performed a number of null and systematics tests to confirm the robustness of our results.

In this paper we explore the modeling of the signal as well as the cosmological consequences of our measurements.  In particular our goal is to measure a combination of the matter density $\Omega_m$ and the amplitude of fluctuations $\sigma_8$ for each tomographic sample.  After checking for consistency, the constraints from all samples can be combined in one measurement with the most statistical power.  

This measurement is particularly timely and interesting: there are hints that the amplitude of lensing at low redshift (which is most sensitive to the parameter $S_8 \equiv \sigma_8 (\Omega_m/0.3)^{0.5}$) from BOSS, KiDS, DES and CFHTLenS (see \cite{Leauthaud:2016jdb, kids1000, Abbott:2017wau, Heymans:2013fya} respectively) might be lower than the most recent constraints from the primary CMB as measured by the Planck satellite \cite{2018arXiv180706209P}. At the same time, hints of a lower-than expected amplitude of low-redshift fluctuations are also present in recent analyses of the redshift space galaxy power spectrum \cite{Ivanov20,DAmico20}.

The combination of CMB lensing and galaxy clustering allows us to test the same physics (gravitational deflection of light low redshift structure) as with galaxy based lensing, but with a very different set of systematics, as our measurement is insensitive to intrinsic alignments, blending and photometric redshift errors of the sources. 
Moreover, if this discrepancy is confirmed, a tomographic measurement with large redshift lever arm such as ours will be essential to investigate its origin. Because our target is measuring low redshift fluctuations only, and to test consistency with galaxy lensing results, we have decided to exclude the CMB lensing power spectrum from the analysis. By only using the unWISE auto spectrum and its cross-correlation with CMB lensing, we are extracting information only over the redshift range of the unWISE sample. The inclusion of the CMB lensing power spectrum would tighten the constraints, but also introduce sensitivity to fluctuations at higher redshifts, which we would like to avoid.

The outline of the paper is as follows: in Section \ref{sec:data} we summarize the properties of the Planck and unWISE data used in this analysis, and in Section \ref{sec:model} we describe the theoretical model used. In Section \ref{sec:mocks}
we present the mocks used to validate our non-linear model and in Section \ref{sec:testmodel} we present the result of these tests on mocks. The cosmological constraints are presented in Section \ref{sec:cosmology}. Finally in Section \ref{sec:conclusions} we summarize our results and directions for future research.

We follow the same convention as in our earlier paper \cite{Krolewski20}, specifically that WISE magnitudes are quoted in the Vega system.  They can be converted to AB magnitudes with AB = Vega + 2.699, 3.339 for the W1 and W2 bands, respectively.

\section{The data}
\label{sec:data}

\begin{figure}
    \centering
    \resizebox{\columnwidth}{!}{\includegraphics{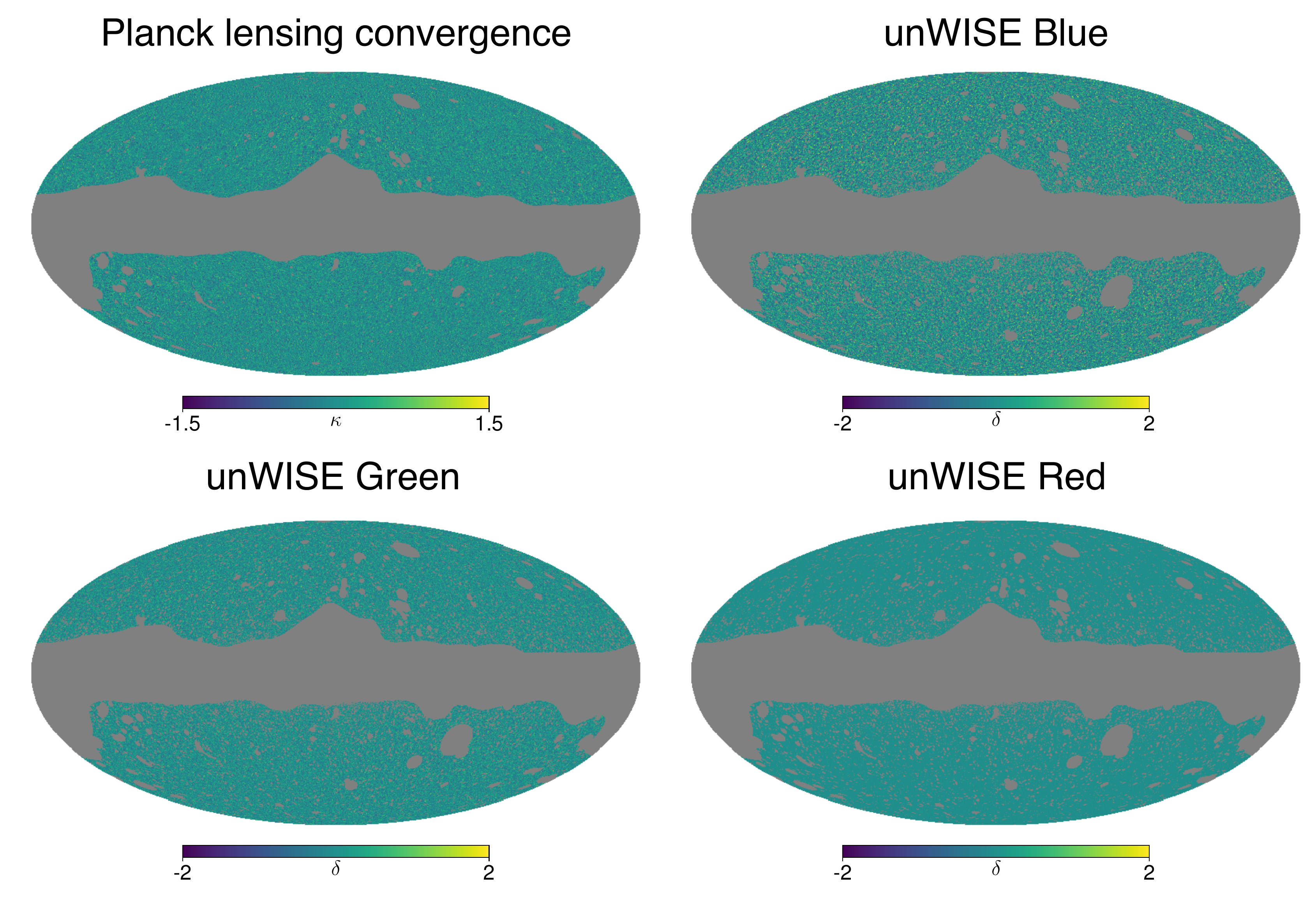}}
    \caption{Plot of the maps used in the analysis ($\kappa$ for Planck lensing convergence and density contrast $\delta$ for the galaxy samples). The maps have been band-pass filtered with \ $\ell_{\rm min} = 100$ and $\ell_{\rm max} = 1000$, and this explains the lack of large-scale power. From ref.~\cite{Krolewski20}.
    }
    \label{fig:sky_maps}
\end{figure}

\subsection{Planck}

Gravitational lensing of the CMB creates a distortion in the temperature and polarization fields that allows for the lensing potential to be reconstructed efficiently (see e.g.\ refs.\ \cite{Lewis06,Hanson10} for reviews).  By searching for these statistical patterns it is possible to reconstruct the lensing convergence, $\kappa$, from quadratic combinations of the foreground-cleaned maps \cite{HuOkamoto02}.
We use the latest CMB lensing maps from the Planck 2018 release \cite{PlanckLens18} and their associated masks, downloaded from the Planck Legacy Archive.\footnote{PLA: \url{https://pla.esac.esa.int/}}
These maps are provided as spherical harmonic coefficients of the convergence, $\kappa_{\ell m}$, in HEALPix format \cite{Gorski05} and with $\ell_{\rm max} = 4096$. For convenience, we convert this to a HEALPix map with $N_{\rm side} = 2048$ by setting $\kappa_{\ell m} = 0$ above $\ell_{\rm max}$.
Our fiducial analysis uses the minimum-variance (MV) estimate obtained from both temperature and polarization, based on the \texttt{SMICA} foreground-reduced CMB map. Since the MV reconstruction is dominated by temperature, residual galactic and extragalactic foregrounds may contaminate the signal. Extensive testing of foreground contamination has been performed by Planck, showing no significant problems at the statistical level of the lensing maps. Since the thermal Sunyaev-Zel'dovich (tSZ) contamination is expected to be one of the largest potential contaminants to cross correlations with tracers of large-scale structure in other analyses \cite{Schaan:2018tup,Madhavacheril:2018bxi,vanEngelen:2013rla,Osborne:2013nna}, we showed (in our previous work \cite{Krolewski20}) that the lensing-galaxy
bandpowers do not shift significantly when using lensing reconstruction on \texttt{SMICA} foreground-reduced maps where tSZ has been explicitly deprojected \cite{PlanckLens18}.


\subsection{unWISE}

We form three galaxy samples using the WISE W1 and W2 magnitudes;
these are the same samples described in ref.~\cite{Krolewski20}, and we refer the reader to that paper for a more comprehensive discussion of them.
Table \ref{tab:samples} gives the adopted color selection for the three samples, which we term the blue, green, and red samples \cite{Schlafly19}.  
Table \ref{tab:samples} also summarizes
important properties of each sample including the redshift distribution,
the number density, galaxy bias, and the response of the number density to galaxy magnification: $s_\mu \equiv d\log_{10}N/dm$ (see discussion in Appendix D of ref.~\cite{Krolewski20}).

\begin{table}[]
\small
\centering
\begin{tabular}{c|ccccccccc}
Label & $\mathrm{W1}-\mathrm{W2} > x$ & $\mathrm{W1}-\mathrm{W2} < x$ & $\mathrm{W2} < x$ & $\bar{z}$ & $\delta z$ & $\bar{n}$ & $b$ & $s_\mu$ &
$k_{\rm max}(\bar{z})$ \\
\hline
Blue & -- & $(17-\mathrm{W2})/4+0.3$ & 16.7 & 0.6 & 0.3 & 3409 & 1.6 & 0.455  & 0.20 \\
Green & $(17-\mathrm{W2})/4 + 0.3$ & $(17-\mathrm{W2})/4 + 0.8$ & 16.7 & 1.1 & 0.5 & 1846 & 2.2 & 0.648  & 0.12 \\
Red & $(17 - \mathrm{W2})/4 + 0.8$ & -- & 16.2 & 1.4 & 0.5 & 144  & 3.3 & 0.842  & 0.09 \\
\end{tabular}
\caption[Summary of unWISE galaxy samples]{Color and magnitude cuts for selecting galaxies of different redshifts, together with the mean redshift ($\bar{z}$) and the standard deviation of the redshift distribution ($\delta z$; as measured by matching to objects with photometric redshifts on the COSMOS field \cite{Laigle16}), number density per deg${}^2$ within the unWISE mask ($\bar{n}$), mean bias ($b$), and response of the number density to magnification ($s_\mu \equiv d\log_{10}N/dm$). 
The final column gives the maximum fitted
wavenumber (in $h$ Mpc$^{-1}$) at $\bar{z}$, i.e.\ $(\ell_{\rm max} + 1/2)/\chi(\bar{z})$.
Galaxies are additionally required to have $\mathrm{W2} > 15.5$, to be undetected or not pointlike in Gaia, and to not be flagged as diffraction spikes, latents or ghosts.
We require that the blue and green samples have $15.5 < \mathrm{W2} < 16.7$, and the red sample has $15.5 < \mathrm{W2} < 16.2$.
See ref.~\cite{Krolewski20} for further details.
}
\label{tab:samples}
\end{table}

As described in ref.~\cite{Krolewski20}, we remove potentially spurious sources and 
each of the samples is required to be either undetected or not pointlike in Gaia,
reducing stellar contamination to $\sim 1\%$.  
The mask is likewise described in ref.~\cite{Krolewski20} and is based on the 2018 Planck lensing mask \cite{PlanckLens18}, with additional cuts to mask bright infrared stars, diffraction spikes, nearby galaxies, and planetary nebulae. The effective sky fraction after masking is $f_{\rm sky} = 0.586$.

\subsubsection{Redshift distribution}

We use two methods for estimating the redshift distribution of our samples, which are described in detail in ref.~\cite{Krolewski20}: (1) Cross-matched redshifts, where unWISE galaxies can be directly matched to galaxies with relatively precise photometric redshifts from the COSMOS photometric catalog \cite{Laigle16} and (2) cross-correlation redshifts with spectroscopic galaxies and quasars from BOSS and eBOSS \cite{Paris12,Reid16,Ata18}. The two distributions are fully consistent with each other \cite{Krolewski20}. 
We set the cross-correlation
redshifts to zero beyond $z = 1.7$ (2.5, 3.0)
for the blue (green, red) samples,
since we find no statistical evidence for any galaxies above this redshift from either the cross-correlation with spectroscopic samples or cross-matched redshifts.
The cross-correlation derived redshifts have considerably smaller statistical uncertainty, and the present analysis will rely primarily on these, because of the added feature that they are a direct measurement of $b(z)\, dN/dz$ rather than $dN/dz$. Given the relatively large width in redshift of these samples and the non-negligible bias evolution with redshift of each sample, a direct measurement of $b(z)\, dN/dz$ is actually desirable, since this product is what enters at lowest order in the modeling of $C_\ell^{gg}$ and $C_\ell^{\kappa g}$ as shown in Section \ref{sec:model}. Some higher order terms, as well as lensing magnification, depend on $dN/dz$ instead, and for those, we use the cross-matched redshifts from COSMOS.  We propagate the uncertainty in $b(z)\, dN/dz$ to uncertainty on cosmological parameters following the procedure described is Section \ref{subsec:dndz_marg}.

\begin{figure}
    \centering
    \resizebox{\linewidth}{!}{\includegraphics{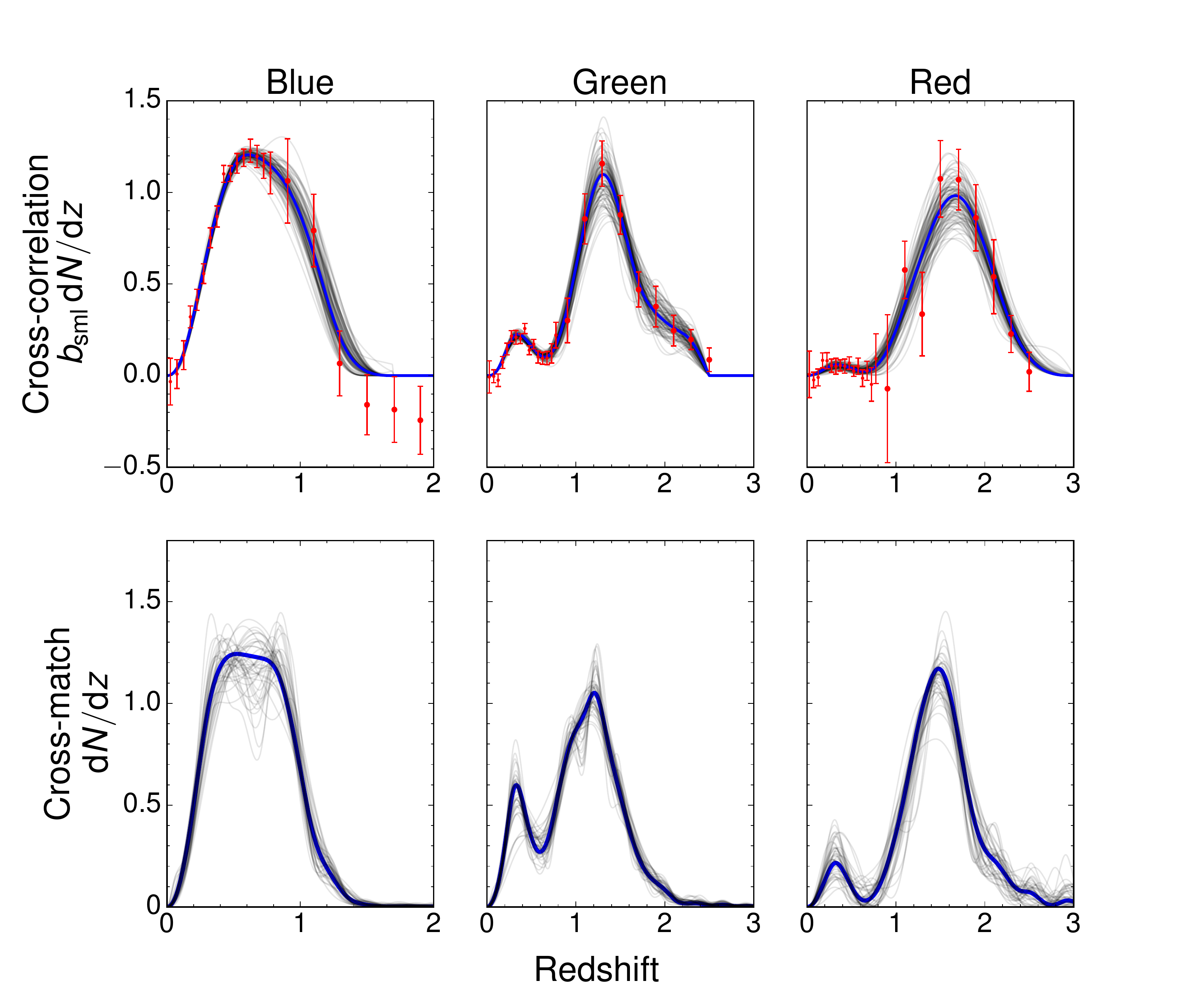}}
    \caption{Redshift distribution derived from cross-correlations with spectroscopic surveys (top row) or photometric redshifts in the COSMOS field (bottom row). Cross-correlation
    redshifts constrain the product of the unWISE galaxy bias and redshift distribution.
    Since $b(z)$ is increasing,
    the cross-correlation $b dN/dz$ is shifted to higher redshifts than the cross-match $dN/dz$.  Red points show the measured
    cross-correlation $b dN/dz$, blue line shows
    the best-fit smooth B-spline, and the light gray lines are samples of redshift
    distributions consistent with the noise in the measured cross-match or cross-correlation redshift distribution. We use these samples
    to propagate the uncertainty in the redshift distribution into our cosmological parameter constraints.
    }
    \label{fig:dNdz}
\end{figure}

\subsection{Angular power spectra}

\begin{figure}
    \centering
    \resizebox{\linewidth}{!}{\includegraphics{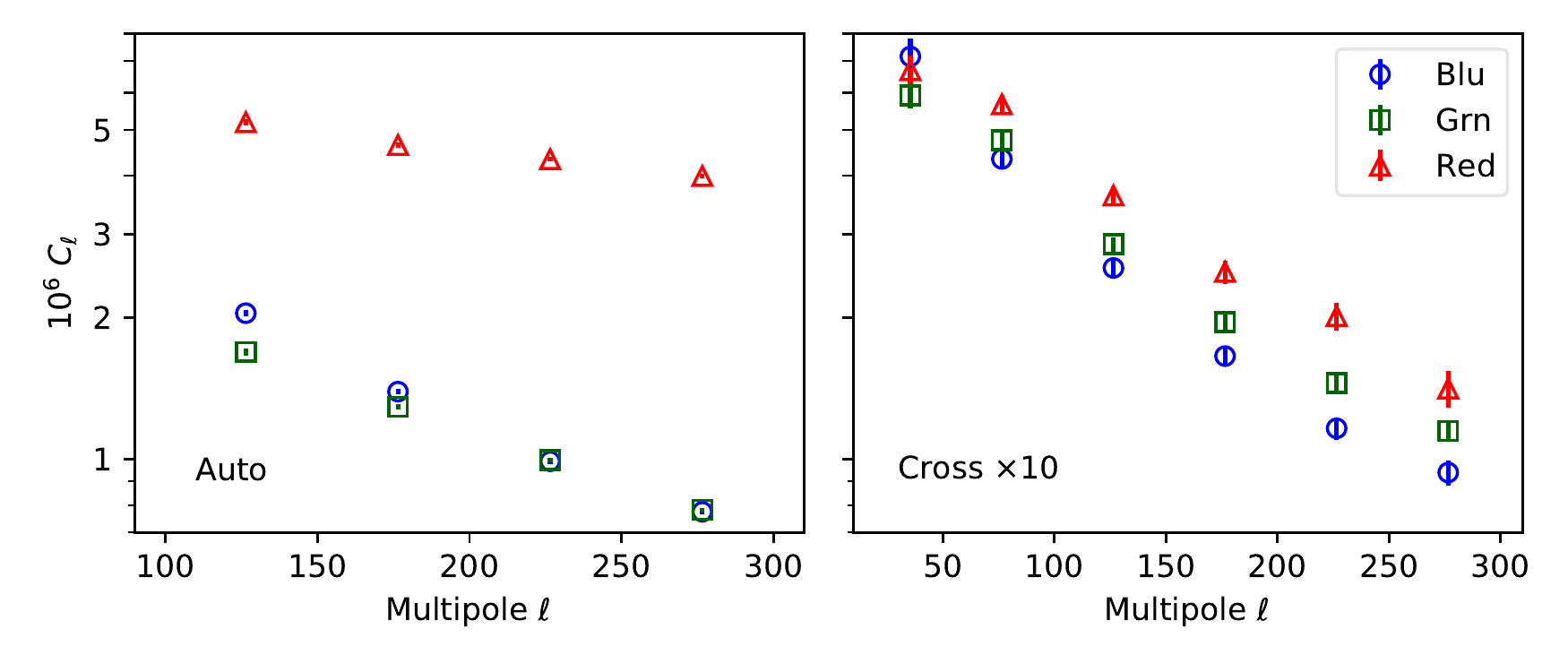}}
    \caption{Angular power spectra for the blue, green and red samples.  The left panel shows the auto spectra ($C_\ell^{gg}$, including shot noise) while the right panel shows the cross-correlation with Planck CMB lensing ($C_\ell^{\kappa g}$; multiplied by 10).  In most cases the statistical error bars are smaller than the symbols.
    }
    \label{fig:data_cl}
\end{figure}

Our measurements consist of the mode-decoupled angular power spectra, $C_\ell^{gg}$ and $C_\ell^{\kappa g}$, for each galaxy sample (Fig.~\ref{fig:data_cl}). Specifically we measure pseudo-$C_\ell$ from the maps (and masks) using the MASTER algorithm \cite{Hivon02} as implemented in the code \texttt{NaMaster}\footnote{\url{https://github.com/LSSTDESC/NaMaster}} \cite{Alonso18}. A full description of the pipeline to obtain the $C_\ell$ is in Section 4.1 of the companion paper \cite{Krolewski20}. In brief, after applying an appropriately apodized mask, the ``mask deconvolved'' or ``decoupled'' $C_\ell$ for both the auto and cross correlations are obtained with \texttt{NaMaster}. The analysis is performed with $\ell_{\rm max}^{\rm NaMaster} = 6000$, to ensure unbiased results, while only the $\ell < 1000$ are retained and used in the analysis and null/systematic tests. We correct for the pixel window function for both the auto and cross-correlations.  We validate our pipeline on a set of 100 simulated Gaussian lensing and galaxy maps (generated with the correct cross-correlation). These simulations have the same power spectrum as the data, including at very low $\ell$, which are not used in our analysis.  Because the unWISE samples have significant amounts of spurious power at $\ell < 20$, we find a few percent biases on large scales due to the mask-induced mode coupling, even after mask decoupling with \texttt{NaMaster}. We suppress these biases by filtering out $\ell < 20$ modes from the unWISE maps before masking, finding that this reduces the biases to $\sim 1\%$.  We further correct for this residual bias with a transfer function (which differs from unity by order a percent) measured on the 100 simulated maps. In conclusion, we have tested that our pipeline can recover both the auto and cross-correlation to sub-percent over the whole range of scales used in the analysis.

We do not use the galaxy
auto-spectra at $\ell < 100$, because in ref.~\cite{Krolewski20}
we find that the low-$\ell$ bandpowers
are contaminated by systematics in the unWISE galaxy
samples.  In contrast,
we find that the galaxy-CMB lensing cross-correlation
is systematics free to $\ell_{\rm min} = 20$,
which is imposed by the fact that we filter
these modes from the unWISE galaxy map.

\subsection{Covariance}

As in our earlier work \cite{Krolewski20}, we approximate the covariance matrix to be given by the disconnected (Gaussian) component only, and further assume that the covariance matrix of the decoupled binned bandpowers of width $\Delta \ell$ is given by \cite{Hivon02}:
\begin{equation}
\label{eq:cov}
    {\rm Cov}(C^{XY}_\ell, C^{XY}_{\ell'}) =  \frac{ \left[ C^{XX}_\ell C^{YY}_\ell + \left(C^{XY}_\ell\right)^2 \right]_{\rm measured} }{f_{\rm sky} (2\ell + 1) \Delta \ell} \ \frac{w_4}{ w_2^2 } \delta_{\ell, \ell'}
\end{equation}
Where $X$ and $Y$ are either $g$ or $\kappa$. Here the weights $w_2$ and $w_4$ are defined in terms of the arbitrary mask weights $W(\hat{\boldsymbol{n}})$ as:
\begin{equation}
    w_i f_{\rm sky} = \frac{1}{4 \pi} \int_{4\pi} d \Omega_{\hat{\boldsymbol{n}}} W^i(\hat{\boldsymbol{n}})
\end{equation}
with $w_1 f_{\rm sky} = f_{\rm sky}$. 

We have also checked that using the more accurate method for analytic Gaussian pseudo-$C_{\ell}$ covariance proposed in \cite{Efstathiou:2003dj, Garcia-Garcia:2019bku}, the largest off-diagonal correlation between bandpowers is 4\% for the two lowest $\ell$ bins, and that the on-diagonal elements agree to percent level.  Therefore we conclude that the approximation in Equation \ref{eq:cov} is adequate for our purposes.
Furthermore, we neglect any non-Gaussian contribution to the covariance matrix, since on the scales analyzed here, those corrections are expected to be negligible.
Finally, we find that our analytic covariance matrix
is very similar to a covariance matrix calculated from the cross-correlation of 300 Planck lensing simulations
with the unWISE galaxy maps.

\section{Model}
\label{sec:model}

Our constraints will be based on measurements of the angular auto-power spectra of the unWISE galaxies $C_\ell^{gg}$ and their cross-spectra with the Planck CMB lensing convergence $C_\ell^{\kappa g}$ (Fig.~\ref{fig:data_cl}).  In what follows we describe our hybrid model for the 3D power spectra of galaxy and matter, and their projection required to predict the auto and cross-correlation between CMB lensing and unWISE galaxies.

\begin{figure}
    \centering
    \includegraphics[width=\columnwidth]{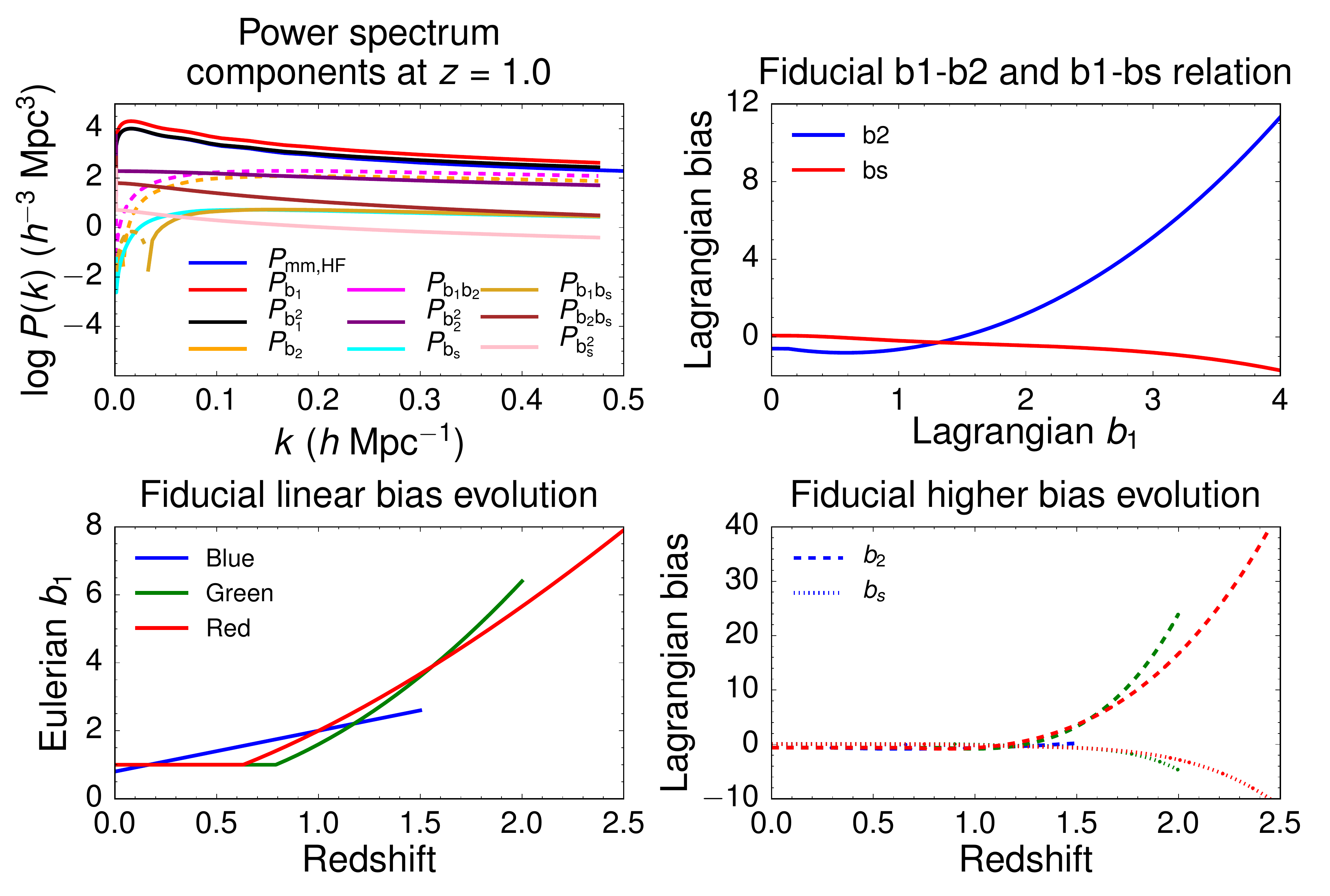}
    \caption{Components of the power spectrum model at $z=1$, fiducial bias relations
    from \cite{Abidi18},
    and fiducial linear bias and higher bias evolution of the unWISE samples. }
    \label{fig:model}
\end{figure}

We model the angular galaxy autocorrelation and galaxy-CMB lensing cross-correlation using a hybrid model that combines fits to N-body simulations with perturbation theory calculations.  Specifically we use a linear bias times the \textsc{HALOFIT} prescription for the matter power spectrum (\cite{Smith03,Takahashi12} as implemented in \texttt{CAMB} \cite{Lewis00,Howlett12}), with additional beyond linear
bias terms from one-loop Lagrangian perturbation theory (specifically Convolution Lagrangian Effective Field Theory \cite{Vlah16,Modi17,Chen20,Chen21}) as implemented in the \texttt{velocileptors} code\footnote{\url{https://github.com/sfschen/velocileptors}}. This hybrid approach correctly models non-linearities in the matter fluctuations, while adding a well-motivated expansion of non-linear bias terms to describe the galaxy field. We validate this approach on realistic mocks of the unWISE samples in Section \ref{sec:mocks} and show that it correctly recovers unbiased cosmological parameters over the range of scales and redshifts considered here.
The matter $m$ and galaxy $g$ power spectra are modeled as: 

\begin{align}
    P_{gm} &= b_{1, E}(z) P_{ mm, HF}(k,z) + \frac{b_{ 2, L}(z)}{2} P_{b_2}(k,z) + \frac{b_{s, L}(z)}{2} P_{b_s}(k,z) 
    \label{eqn:pmg}
    \\
    P_{gg} &= b_{1, E}(z)^2 P_{mm, HF}(k,z) + b_{2, L}(z) P_{b_2}(k,z) + b_{s, L}(z) P_{b_s}(k,z) \nonumber \\
    &\qquad+ b_{1, L}(z) b_{2, L}(z) P_{b_1 b_2}(k,z)
    + b_{1, L}(z) b_{s, L}(z) P_{b_1 b_s}(k,z)
    + b_{2, L}(z) b_{s, L}(z) P_{b_2 b_s}(k,z) \nonumber \\
    &\qquad+ b_{2, L}(z)^2 P_{b_2^2}(k,z)
    + b_{s, L}(z)^2 P_{b_s^2}(k,z) \nonumber \\
    &\qquad + \textrm{Shot Noise} 
    \label{eqn:pgg}
    \\
    P_{mm} &= P_{mm, HF}(k,z)
\end{align}
where $P_{mm, HF}$ is the HALOFIT matter power spectrum 
of non-neutrino density fluctuations
and the subscripts $E$ and $L$ indicate the Eulerian and Lagrangian biases respectively,
with $b_{1,E} = b_{1,L} + 1$.
The power spectrum contributions $P_{b_2}$, $P_{b_s}$, etc.\
are computed analytically as a function of the linear power spectrum $P_{\rm lin}(k,z)$
following Equation 3.1 in ref.~\cite{Chen21}
and shown in Fig.~\ref{fig:model} at $z = 1$.
The shot-noise is scale-independent, i.e.\ a constant.

This is equivalent to making the following substitutions
in the CLEFT equations for $P_{gg}$ and $P_{gm}$
(as given in ref.\ \cite{Kitanidis21}, originally equation 2.7 of ref.\ \cite{Modi17} and equation B.2 of ref.~\cite{Vlah16}; similar
to the approach in ref.~\cite{Pandey20}
for modelling cosmic shear)
\begin{equation}
    \left(1 - \frac{\alpha_{\rm cross} k^2}{2}\right) P_{Z} + P_{\rm 1-loop} \rightarrow P_{mm, HF}
\end{equation}
\begin{equation}
    \left(1 - \frac{\alpha_{\rm auto} k^2}{2}\right) P_{Z} + P_{\rm 1-loop} \rightarrow P_{mm, HF}
\end{equation}
\begin{equation}
    P_{b_1} \rightarrow 2 P_{mm, HF}
\end{equation}
\begin{equation}
    P_{b_1^2} \rightarrow P_{mm, HF}
\end{equation}
where $P_{Z}$ is the Zeldovich contribution,
$P_{\rm 1-loop}$ is the 1-loop contribution,
and $\alpha_{\rm cross}$ and $\alpha_{\rm auto}$
are EFT parameters encapsulating small-scale
physics that cannot be modelled by perturbation theory.
In ref.\ \cite{Modi17} the EFT counterterm $\alpha$
was allowed to differ between $P_{gg}$ and $P_{gm}$
to encompass other terms neglected in that analysis.
Thus the first two substitutions replace
the perturbative plus EFT expansion with the empirical
matter power spectrum,
whereas the third and fourth slightly simplify the bias power spectra.
Finally, while $P_{mm, HF}(k,z)$
varies with the cosmological parameters,
we fix the higher-order terms $P_{b_2}(k,z)$, $P_{b_s}(k,z)$, etc.\ to their values at the fiducial cosmology (the true cosmology of the simulation,
or the Planck 2018 cosmology in data).
 
We use the Limber approximation
\cite{Limber53,Loverde08}
to project the 3D galaxy power spectra
into angular power spectra:
\begin{equation}
    C_{\ell}^{gg} = \int d\chi \frac{W^{g}(\chi)^2}{\chi^2} P_{gg}(k\chi = \ell + 1/2, z)
\label{eqn:limber_gg}
\end{equation}
\begin{equation}
    C_{\ell}^{\kappa g} = \int d\chi \frac{W^{g}(\chi)W^{\kappa}(\chi)}{\chi^2} P_{gm}(k\chi = \ell + 1/2, z)
\label{eqn:limber_kg}
\end{equation}
where the galaxy kernel and CMB lensing kernels are
\begin{equation}
    W^g(\chi) = \frac{dN}{d\chi}
    \quad , \quad
    W^{\kappa}(\chi) = \frac{3}{2} \Omega_{m}  H_0^2 (1+z) \frac{\chi (\chi_{\star} - \chi)}{\chi_{\star}}
\end{equation}
where $\Omega_m$ is the low-redshift
matter density, i.e.\ including the neutrino
density $\Omega_{\nu}$,
and $\chi_\star$ the comoving distance to the surface of last scattering at $z \approx 1100$. Note that since we include bias evolution
in $P_{gm}$ and $P_{gg}$ (Equations~\ref{eqn:pmg} and~\ref{eqn:pgg}),
we do not include bias terms in the galaxy kernel $W^g$.  

We also include magnification bias in our
modelling, as it contributes
to the angular power spectra at the few percent level \cite{Krolewski20}. The magnification bias kernel
for a galaxy sample $i$ \cite{Villumsen95, Moessner98,BartelmanSchneider01}
is given by
\begin{equation}
W^{\mu,i}(\chi) =  (5s_\mu-2)\,\frac{3}{2} \Omega_m  H_0^2(1+z) g_i(\chi)
\end{equation}
\begin{equation}
g_i(\chi) = \int_{\chi}^{\chi_{\star}} d\chi' \ \frac{\chi(\chi' - \chi)}{\chi'} \ H(z') \ \frac{dN_i}{dz'}
\end{equation}
where $s_\mu$ is the response of the galaxy
number density to a change in magnitude (Table \ref{tab:samples}).
Therefore our total model 
for the observed angular power spectra
is given by
\begin{equation}
    C_{\ell}^{\rm unWISE, unWISE} = C_{\ell}^{gg} + 2 C_{\ell}^{g \mu} + C_{\ell}^{\mu \mu}
\end{equation}
\begin{equation}
    C_{\ell}^{\rm \kappa, unWISE} = C_{\ell}^{\kappa g} + C_{\ell}^{\kappa \mu}
\end{equation}

The Limber approximation is accurate to $<1\%$ on the scales that we consider ($\ell > 20$) \cite{Hu00,Verde00,Loverde08} so we do not consider extensions beyond the first-order correction: $\ell \rightarrow \ell + 1/2$ \cite{Loverde08}.

In our fiducial model and measurement,
we fix the higher-bias evolution
$b_{2,L}(z)$ and $b_{s,L}(z)$ (Fig.~\ref{fig:model})
and use cross-correlation redshifts that estimate the
product of $b_{1, E}(z)$ and the redshift kernel.
We use cross-correlation redshifts
rather than cross-match redshifts in the leading-order term
both to allow us to include $b_{1, E}(z)$
and because the uncertainties on the cross-correlation
redshifts are smaller and better understood.
For higher-order terms, we use cross-match redshifts
from the COSMOS field
to directly estimate $dN/d\chi$.

The evolution of $b_{2,L}$ and $b_{s,L}$ (Fig.~\ref{fig:model})
is determined from the relations
between $b_{1,L}$, $b_{2,L}$ and $b_{s,L}$ presented in Fig.\ 8 of
\cite{Abidi18}, as directly
measured from the clustering of protohalos in $N$-body simulations.
This agrees very well
with separate-universe simulations \cite{Lazeyras15}
or local Lagrangian biasing
within the Sheth-Tormen or Press-Schechter
mass function and we assume that these halo relationships are
indicative of those of the galaxies
(e.g.\ as seen in \cite{Barreira21}).

In the data, for the fiducial $b_{1,L}(z)$ evolution
required for the fiducial $b_{2,L}(z)$
and $b_{s,L}(z)$, we use \cite{Krolewski20}
\begin{subequations}
\begin{align}
   b_{1,L,{\rm fid}}(z) &= 0.8 + 1.2 z - 1 &&\text{Blue} \\
   b_{1,L,{\rm fid}}(z) &= \max{(1.6z^{2},1)} - 1 &&\text{Green} \\
   b_{1,L,{\rm fid}}(z) &= \max{(2z^{1.5},1)} - 1 &&\text{Red}
\end{align}
\label{eqn:bias_evolution}
\end{subequations}
with $\max{(a,b)}$ meaning the larger of $a$
and $b$.

In the simulations,
we directly measure the bias evolution
of the simulated galaxies and use this
for $b_{1,L}(z)$ (Fig.~\ref{fig:bias_comparison}).

On large scales cross-correlation redshifts measure
the product of the redshift
distribution and the linear Eulerian bias evolution:
\begin{equation}
    \hat{W}^{{\rm xc}, g}(\chi) \propto b_{1, E}(z) \frac{dN}{d\chi}
\end{equation}
We note that in principle the equation above has some (weak) dependence on the fiducial cosmology assumed when obtain cross-correlation redshifts. On the contrary, no fiducial cosmology needs to be assumed to obtain direct matching redshifts from COSMOS.
The normalization of cross-correlation redshifts such that $\int d\chi\ \hat{W}^{{\rm xc}, g}(\chi) \equiv 1$ partly removes the dependence on the fiducial cosmology, by shifting that dependence to the galaxy bias (which is marginalized over in our constraints).

In our fiducial analysis of the data, we fix the distance-redshift relation (i.e.\ explicit factors of $\chi(z)$ and $H(z)$) appearing in Eqs.~\ref{eqn:limber_gg} and~\ref{eqn:limber_kg}, to that in our fiducial cosmology.
We refer to this procedure as ``fixed geometry'', and whenever analyzing the real data, we use Planck 2018 cosmological parameters as fiducial. To validate this procedure, after the cosmological fits we updated the cosmology to the best-fit cosmology obtained from the green sample (the most constraining) and repeat the analysis. The cosmological parameters shift by $<0.2 \sigma$ when updating the fiducial cosmology in the redshift-distance relation, and therefore we conclude that our procedure is appropriate given our statistical uncertainty.

For terms that require $dN/dz$ rather than $b(z)dN/dz$ (such as magnification or higher order biases), we note that $dN/dz$ is the quantity directly measured by cross-matching with COSMOS and therefore we fix $dN/dz$ and allow $H(z)$ to vary with the cosmological parameters.

In tests on mocks, we test both the ``fixed geometry'' procedure, as well as a ``free geometry'' one in which we fix $b(z)\, dN/dz$, and allow $H(z)$ to vary. We find very similar constraints as shown in Table~\ref{tab:green_test}. 

If we define $b_{1,E}^{\rm eff}$ as the weighted average of $b_{1,E}(z)$ over the redshift distribution:
\begin{equation}
    b_{1,E}^{\rm eff} \equiv \int d\chi \ b_{1,E}(z) \frac{dN}{d\chi}
\label{eqn:b1_def}
\end{equation}
we may write
\begin{equation}
    \hat{W}^{{\rm xc}, g}(\chi) = \frac{b_{1,E}(z) dN/d\chi}{b_{1,E}^{\rm eff}}
\label{eqn:what_def}
\end{equation}
Integrating Equation~\ref{eqn:what_def} over $\chi$ gives unity as required.

Substituting the sum in Equation~\ref{eqn:pgg}
into Equation~\ref{eqn:limber_gg}
yields a sum
of angular power spectra
\begin{equation}
    C_{\ell}^{gg} =  C_{\ell}^{gg, (I)}
    +C_{\ell}^{gg, (II)}
    +C_{\ell}^{gg, (III)}
    +C_{\ell}^{gg, (IV)}
    +C_{\ell}^{gg, (V)}
    +C_{\ell}^{gg, (VI)}
    +C_{\ell}^{gg, (VII)}
    +C_{\ell}^{gg, (VIII)}
\end{equation}
For $C_{\ell}^{gg, (I)}$,
we can use
Equation~\ref{eqn:what_def} to replace 
$dN/d\chi$ with the observable $\hat{W}^{{\rm xc}, g}(\chi)$, yielding
\begin{equation}
C_{\ell}^{gg, (I)} = \int d\chi \left(b_{1,E}^{\rm eff}\right)^2 \frac{\hat{W}^{{\rm xc},g}(\chi)^2}{\chi^2} P_{mm, HF}(k,z)
\label{eqn:cell_one}
\end{equation}
The second term requires the direct
measurement of $dN/dz$ from the COSMOS field,
$\hat{W}^{{\rm dir},g}(z)$
\begin{equation}
 C_{\ell}^{gg, (II)} = \int d\chi \frac{\hat{W}^{{\rm dir},g}(z)^2 H(z)^2}{N^2 \chi^2} b_{2,L}(z) P_{b_2}(k,z)
\end{equation}
where $N$ is the normalization
\begin{equation}
N \equiv \int d\chi \hat{W}^{{\rm dir},g}(z)^2 H(z)
\end{equation}
and $b_{2,L}$ is a function of redshift through its dependence on $b_{1,L,{\rm fid}}(z)$.
Likewise for the third term
\begin{equation}
 C_{\ell}^{gg, (III)} = \int d\chi \frac{\hat{W}^{{\rm dir},g}(z)^2 H(z)^2}{N^2 \chi^2} b_{s,L}(z) P_{b_s}(k,z)
\end{equation}
The cross terms with $b_{1,L}$ require both $\hat{W}^{{\rm dir},g}(z)$ and $\hat{W}^{{\rm xc},g}(\chi)$ and recalling $b_{1,E}=1+b_{1,L}$ we see that
\begin{align}
 C_{\ell}^{gg, (IV)} &= b_{1,E}^{\rm eff} \int d\chi \frac{\hat{W}^{{\rm dir},g}(z) H(z) \hat{W}^{{\rm xc},g}(\chi)}{N \chi^2} b_{2,L}(z) P_{b_1 b_2}(k,z) \nonumber \\
 &\quad - \int d\chi \frac{\hat{W}^{{\rm dir},g}(z)^2 H(z)^2}{N^2 \chi^2} b_{2,L}(z) P_{b_1 b_2}(k,z)
\end{align}
\begin{align}
 C_{\ell}^{gg, (V)} &= b_{1,E}^{\rm eff} \int d\chi \frac{\hat{W}^{{\rm dir},g}(z) H(z) \hat{W}^{{\rm xc},g}(\chi)}{N \chi^2} b_{s,L}(z) P_{b_1 b_s}(k,z) \nonumber \\
 &\quad - \int d\chi \frac{\hat{W}^{{\rm dir},g}(z)^2 H(z)^2}{N^2 \chi^2} b_{s,L}(z) P_{b_1 b_s}(k,z)
\end{align}
The final three terms require only $\hat{W}^{{\rm dir},g}(z)$ 
\begin{equation}
 C_{\ell}^{gg, (VI)} = \int d\chi \frac{\hat{W}^{{\rm dir},g}(z)^2 H(z)^2}{N^2 \chi^2} b_{2,L}(z) b_{s,L}(z) P_{b_2 b_s}(k,z)
\end{equation}
\begin{equation}
 C_{\ell}^{gg, (VII)}= \int d\chi \frac{\hat{W}^{{\rm dir},g}(z)^2 H(z)^2}{N^2 \chi^2} b_{2,L}(z)^2 P_{b_2^2}(k,z)
\end{equation}
\begin{equation}
 C_{\ell}^{gg, (VIII)} = \int d\chi \frac{\hat{W}^{{\rm dir},g}(z)^2 H(z)^2}{N^2 \chi^2} b_{s,L}(z)^2 P_{b_s^2}(k,z)
\end{equation}
$C_{\ell}^{\kappa g}$ does not require any mixed
terms
\begin{align}
 C_{\ell}^{\kappa g} &= \int d\chi b_{1,E}^{\rm eff} \frac{\hat{W}^{{\rm xc},g}(\chi)W^{\kappa}(\chi)}{\chi^2} P_{mm, HF}(k,z) \nonumber \\
 &+ \int d\chi \frac{\hat{W}^{{\rm dir},g}(z) H(z) W^{\kappa}(\chi)}{N \chi^2} \frac{b_{2,L}(z)}{2} P_{b_2}(k,z) \nonumber \\
 &+ \int d\chi \frac{\hat{W}^{{\rm dir},g}(z) H(z) W^{\kappa}(\chi)}{N \chi^2} \frac{b_{s,L}(z)}{2} P_{b_s}(k,z)
 \label{eqn:cell_kg}
\end{align}
For $C_{\ell}^{\kappa \mu}$
and $C_{\ell}^{\kappa \mu}$,
we use $\hat{W}^{{\rm dir},g}(z)$ 
in the galaxy kernel,
and for $C_{\ell}^{g \mu}$,
we use $\hat{W}^{{\rm xc},g}(\chi)$.

The combination of the cross-correlation and cross-match redshifts allows us to perform a consistency check on the higher-order terms in our model.  When integrating over comoving distance in Equations~\ref{eqn:cell_one} to~\ref{eqn:cell_kg}, we assume
that the evolution of $b_1$ is proportional to the ratio between cross-correlation
and cross-match redshifts. However, the small-scale clustering
used for measuring the cross-match redshifts will also be sensitive to the higher-bias terms.
This contribution cannot be so large that it changes
the effective small-scale bias (i.e.\ ratio of the galaxy power spectrum in our model, including higher-order bias contributions, to the matter power spectrum) to be much different from the linear bias.
To assess self-consistency, we divide the cross-correlation redshift distribution by the cross-match redshift distribution and multiply by the best-fit linear bias from Section~\ref{sec:results} and Fig.~\ref{fig:big_corner_plot} to get $b_1(z)$. Then we evaluate Equation~\ref{eqn:pgg} for $P_{gg}(z)$, Fourier transform to the correlation function, integrate to get the projected correlation function $w_p$, and average over the relevant scales (2.5 to 10 $h^{-1}$ Mpc), as in Equation~\ref{eqn:bsml}. We find that the difference between this effective small-scale bias and $b_1(z)$ is much smaller than the measurement error on the bias from uncertainty in the clustering measurement (i.e.\ errorbars in Fig.~\ref{fig:bias_comparison}).  We therefore conclude that our model is self-consistent, within the errorbars on the spectroscopic cross-correlation.

In our fiducial model, we vary $b_{1,E}^{\rm eff}$, $s_\mu$
and the shot
noise for each sample, along with the two
cosmological parameters, $\Omega_m$ and $\sigma_8$.  We hold the higher-order biases fixed and only evaluate the higher bias
power spectra $P_{b_2}$, $P_{b_s}$, etc.\ at the fiducial cosmology.
We tested models in which we allowed
an overall scaling in $b_{2,L}(b_{\rm 1, {\rm fid}}(z))$
or added a constant to $b_{2,L}(b_{\rm 1, {\rm fid}}(z))$
but these did not change our results.

\section{Mocks}
\label{sec:mocks}

In this section we describe the mocks that we use
to test our cosmological inference pipeline.
These mocks are intended to create a plausible
sample of mock galaxies to test the impact of non-linearities,
scale-dependent bias, and uncertain redshift distribution,
rather than representing a faithful HOD model
of the unWISE galaxies.  Since our goal is to \emph{test}
our pipeline, rather than \emph{calibrate} any aspect of our
model or covariance we primarily need a sample of objects
with similar clustering and redshift distribution to our data
with a plausibly complex relationship to the underlying matter
distribution.

\begin{figure}
    \centering
    \includegraphics[width=0.4\textwidth]{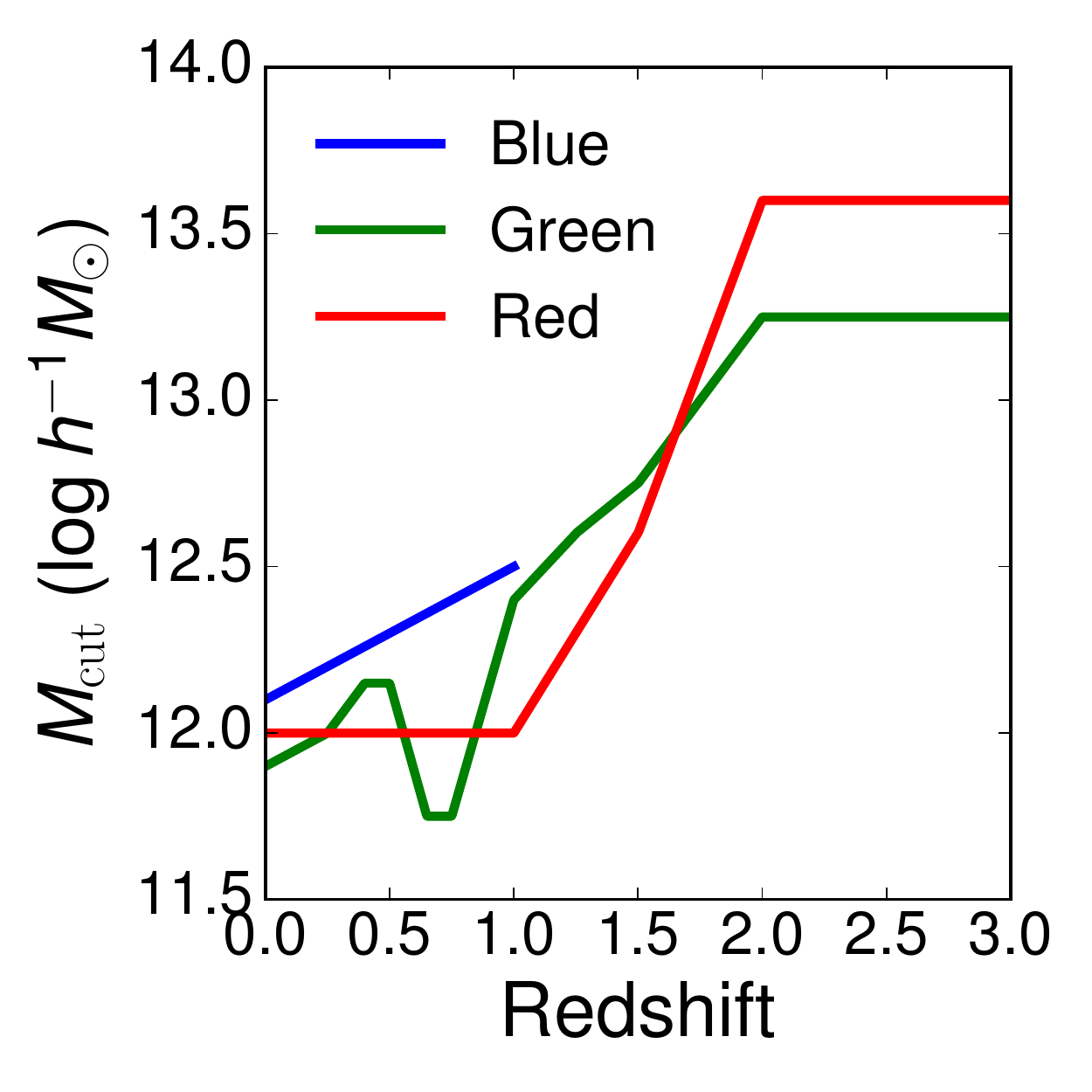}
    \includegraphics[width=0.4\textwidth]{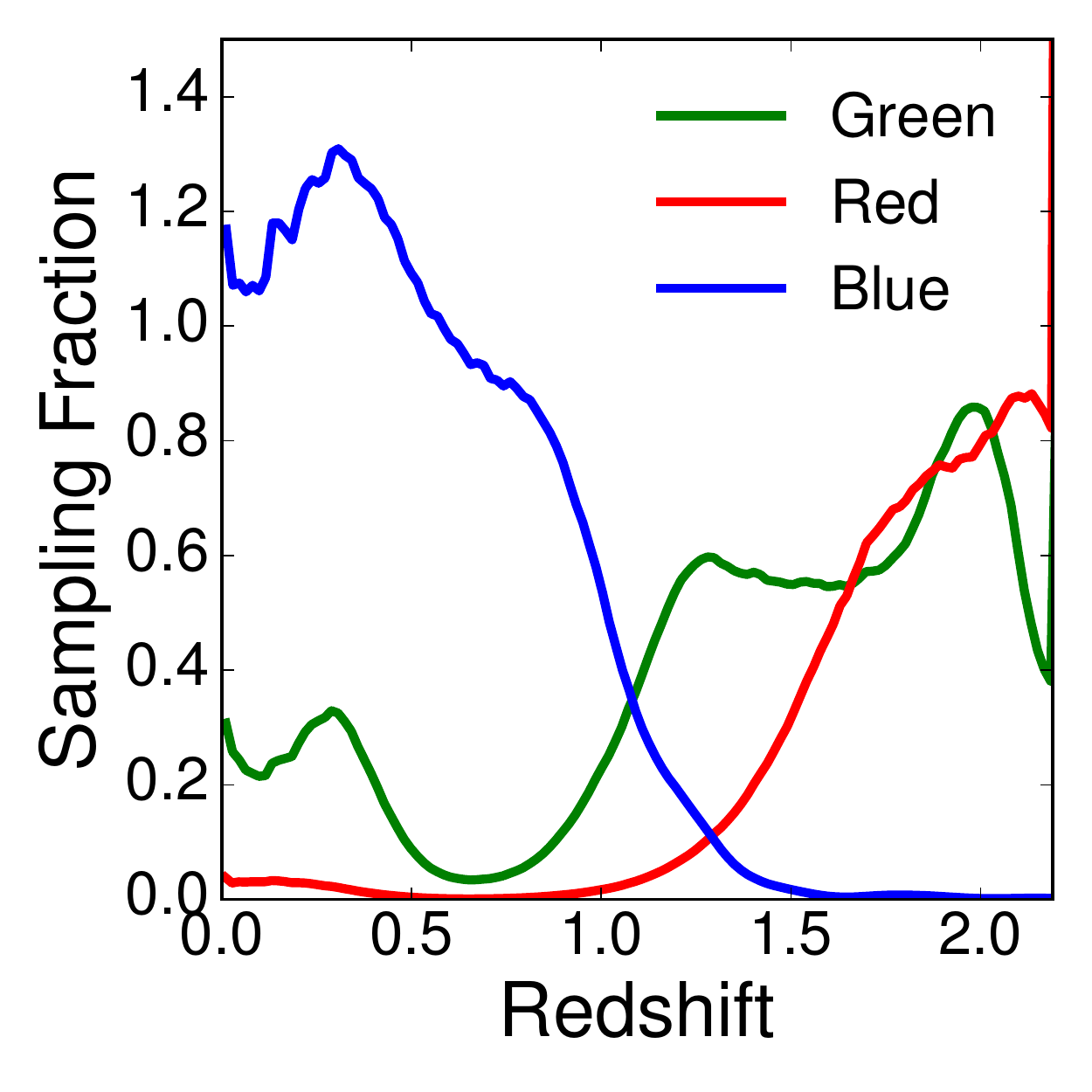}
    \caption{\textit{Left:} $M_{\rm cut}$ as a function of
    redshift for the HOD of the 3 unWISE
    mock galaxy samples. \textit{Right:} Sampling fraction (i.e.\ factor by which the HOD of Eqs.~\ref{eqn:hod_cen} and~\ref{eqn:hod_sat} is downsampled) as a function of redshift.
    When the sampling fraction exceeds unity, we augment the catalog with a Poisson resampling of the halo catalog
    to better match the required $dN/dz$.}
    \label{fig:mcut_vs_redshift}
\end{figure}

\begin{figure}
    \centering
    \includegraphics[width=\textwidth]{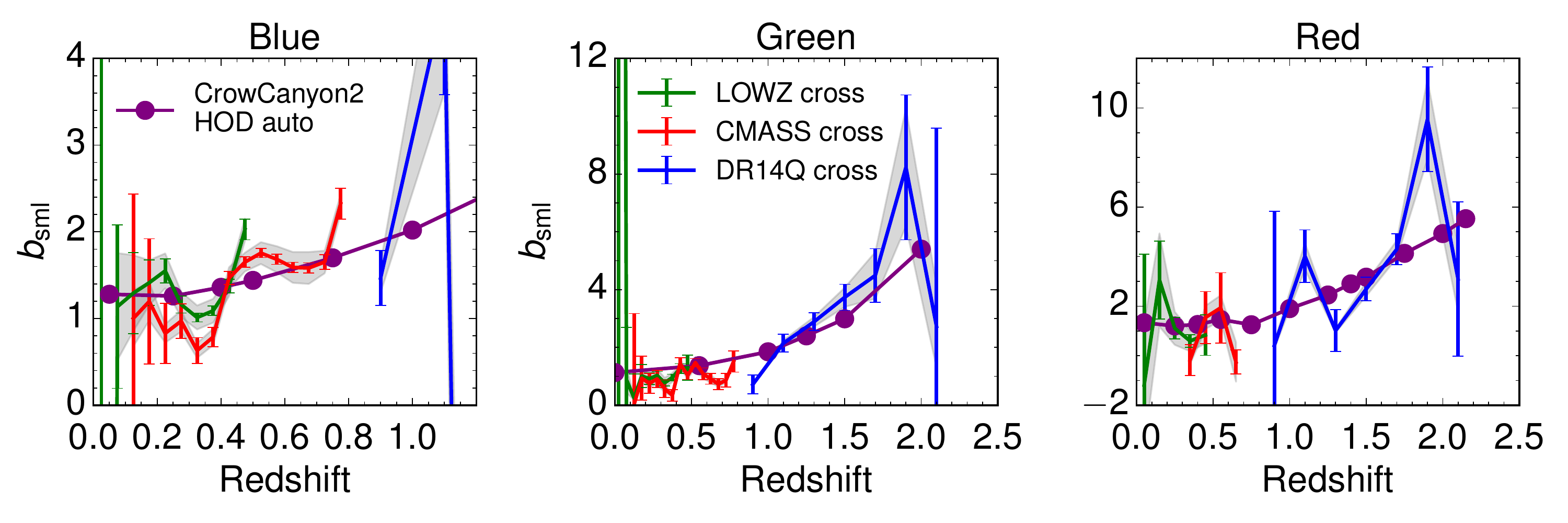}
    \caption{Comparison between bias for HODs in the CrowCanyon2 simulation (purple) and measured bias from cross-correlations with LOWZ (green), CMASS (red), and DR14 quasars (blue). }
    \label{fig:bias_comparison}
\end{figure}

\begin{figure}
    \centering
    \includegraphics[width=\textwidth]{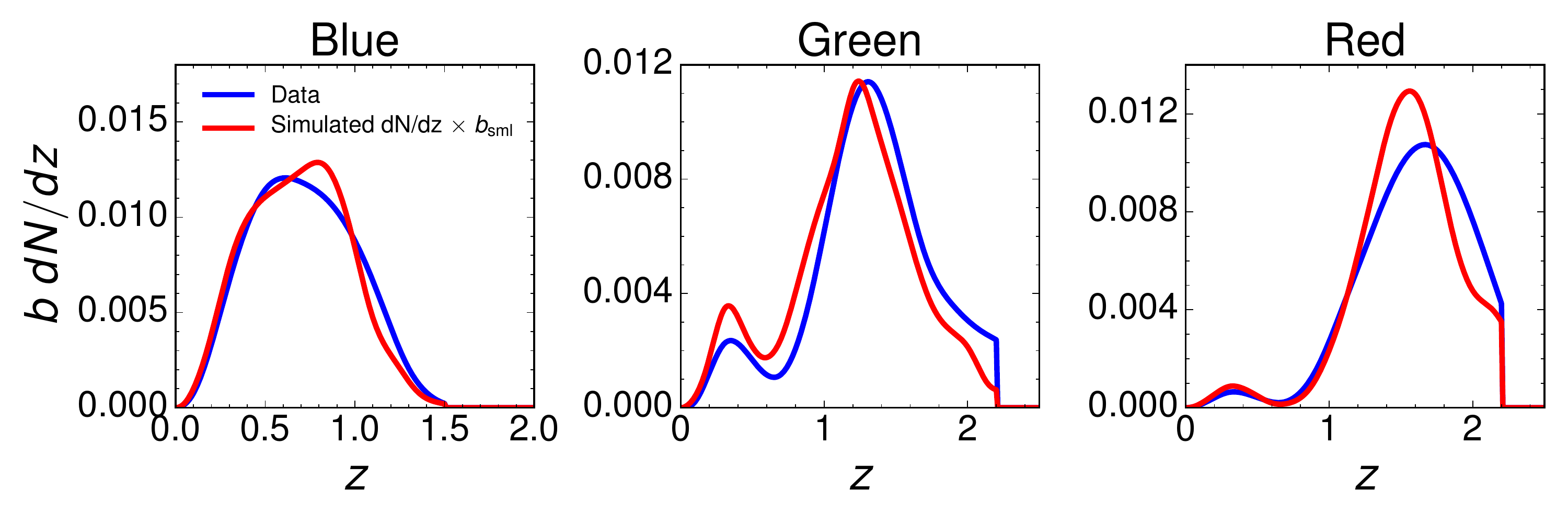}
    \caption{Comparison between cross-correlation redshift measurement of $b\, dN/dz$, and the $b_{\rm sml}$ measurement in Fig.~\ref{fig:bias_comparison} multiplied by the cross-match redshift distribution. 
    The data redshift distributions are truncated
    at $z = 2.2$, the maximum
    redshift of the CrowCanyon2 simulation.}
    \label{fig:dndz_comparison}
\end{figure}

To this end we model the unWISE galaxies using a simple HOD applied to dark matter halos in an $N$-body simulation. We simply use a 6-parameter family\footnote{There is some evidence that HOD parameters scale approximately universally with number density, e.g.~ref.~\cite{Brown08}.  A similar assumption is at the root of the `SHAM' approximation \cite{SHAM}.} of HODs based on ref.~\cite{Zheng05} with 
\begin{equation}
  \langle N_{\rm cen} \rangle = \frac{1}{2}f_{\rm samp}\left[1 + {\rm erf}\left(\frac{{\rm log_{10}} M - {\rm log_{10}} M_{\rm cut}}{\sqrt{2}\sigma_{{\rm log_{10}}M}}\right)\right]
  \qquad 
\label{eqn:hod_cen}
\end{equation}
and
\begin{equation}
  \langle N_{\rm sat} \rangle = \left[ \frac{M - \kappa M_{\rm cut}}{\beta M_{\rm cut}}\right]^{\alpha}
  \quad .
\label{eqn:hod_sat}
\end{equation}
We adjust the parameters by hand to match
the number density (Fig.~\ref{fig:mcut_vs_redshift}), bias evolution (Fig.~\ref{fig:bias_comparison})
and therefore the bias-weighted
redshift distribution
constrained by the clustering redshifts
(Fig.~\ref{fig:dndz_comparison}).
Specifically,
we keep $\sigma$, $\kappa$, $\alpha$,
and $\beta$ constant, adjust
$M_{\rm cut}$ with redshift
to roughly match the bias evolution
(Fig.~\ref{fig:bias_comparison}),
and exactly adjust $f_{\rm samp}$
to match the redshift distribution.
If $f_{\rm samp} > 1$,
we Poisson re-sample the halo catalog
(i.e.\ some halos are double-counted)
to exactly match $dN/dz$.
We favor $f_{\rm samp} < 1$,
but find $f_{\rm samp} > 1$ is
required for the less massive
blue sample due to the mass resolution
of the simulation.

The requirement of matching
both the number density
and the bias evolution leads
to features such as the dip in $M_{\rm cut}$ at $z \sim 0.7$ in green,
to match a drop in the bias in data,
which also requires a drop in $f_{\rm samp}$ to match the number density.
Such features may indicate the limitations of our modeling; however we
require primarily that the clustering and number density are similar to
that of the unWISE data and not that the model be physically compelling.
Recall these mocks are not used to calculate the theoretical predictions
or covariance matrices. 

For blue, we use $\kappa = 0.1$, 
$\alpha = 1.0$ and $\beta = 15$.
We find that a redshift-dependent
$\sigma$ works best, with $\sigma = 0.22$
at $z > 0.5$, 0.43 at $z < 0.4$, and a linear
ramp from $0.4 < z < 0.5$.
For green, we use
$\sigma = 0.25$, $\kappa = 0.1$, $\alpha = 1.0$, and 
$\beta = 10$; and
for red, we use $\sigma = 0.43$, $\kappa = 1.0$, 
$\alpha = 0.8$ and $\beta = 15$.
We plot $M_{\rm cut}$ and $f_{\rm samp}$ as a function of 
redshift in Fig.~\ref{fig:mcut_vs_redshift}.

We populate halos from one of the simulations in the CrowCanyon2
simulation suite (see Appendix \ref{app:crowcanyon2}), specifically
a \texttt{FastPM} \cite{FengChuEtAl16} $N$-body simulation
with $8192^3$ particles in a $4096$ $h^{-1}$Mpc box.  The force
resolution factor $B$ is 3 and 40 time steps
are used between $z = 19$ and $z = 0$.
We use the full-sky lightcone output
from CrowCanyon2, and halos
are identified using a friends-of-friends halo
finder with linking length $b = 0.2$.
The simulation
uses a $\Lambda$CDM cosmology
close to the Planck 2018 cosmology,
with $\Omega_m = 0.3092$, $\Omega_b = 0.0496$,
$h = 0.677$, $n_s = 0.968$, and $\sigma_8 = 0.822$. While the galaxy map is obtained by populating Dark Matter halos with the HOD described above, the CMB lensing maps are obtained by using the Born approximation\footnote{The Born approximation is expected to be an excellent approximation on the scales of interest and for Planck noise level \cite{Pratten:2016dsm, Marozzi:2016uob, Boehm:2019hlv, Fabbian:2019tik}.} and integrating the matter density on the lightcone, weighted by the CMB lensing kernel, as described in Appendix \ref{app:crowcanyon2}.
The simulated volume is quite large ($69\,h^{-3}\mathrm{Gpc}^3$) but not overwhelmingly larger than the volume sampled by the data.  Since much of our constraining power comes from large scales, we expect some fluctuations in the galaxy clustering due to the sample variance from the initial conditions.

To make sure that the HOD roughly reproduces the correct bias evolution as seen in the data, we compare to observations of cross-clustering with
spectroscopic galaxies (Fig.~\ref{fig:bias_comparison}) by considering the quantity 
\begin{equation}
b_{\rm sml} = \sqrt{\bar{w}_{\rm gg}/\bar{w}_{\rm mm}}
\quad\mathrm{with}\quad
\bar{w} = \int_{r_{p, \rm min}}^{r_{p, \rm max}} dr_p\ w_p(r_p)
\label{eqn:bsml}
\end{equation}
where $r_{p, \rm min} = 2.5$ $h^{-1}$ Mpc and $r_{p, \rm max} = 10$ $h^{-1}$ Mpc.
This matches the scales on which the cross-correlation redshifts are measured.

\begin{figure}
    \centering
    \includegraphics[width=\textwidth]{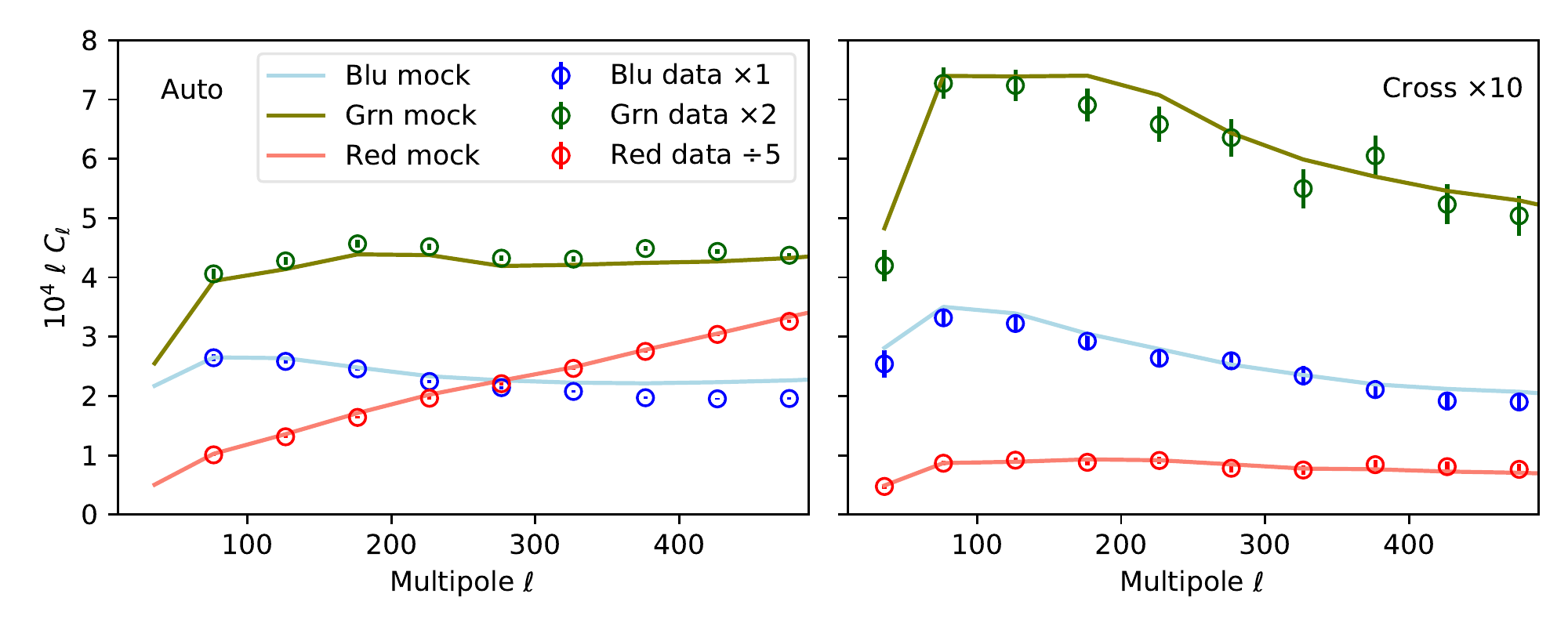}
    \caption{Comparison between $C_{\ell}^{gg}$ (left) and $C_{\ell}^{\kappa g}$ (right) in the data (points with error bars) and in the CrowCanyon2 simulation (lines) for the blue, green and red samples.  To bring them to a common scale while separating the lines for readability we have multiplied the green sample values by $2$ and divided the red sample values by $5$.  The cross spectra have all been multiplied by an additional factor of $10$ compared to the auto-spectra.
    }
    \label{fig:cl_comparison}
\end{figure}

The comparison of $b(z)dN/dz$ between the mocks and the data is shown in Fig. \ref{fig:dndz_comparison}, while the comparison of $C_{\ell}^{\kappa g}$ and $C_{\ell}^{gg}$ on the lightcone is shown in Fig.~\ref{fig:cl_comparison}. We find qualitative agreement for the bias evolution, redshift distribution and auto/cross-power spectra. Note that the purpose of these mocks is to test that our non-linear model is flexible enough to recover unbiased parameters from the underlying simulation, so that only qualitative agreement with data is required.

\section{Testing the model on mocks}
\label{sec:testmodel}

\subsection{Recovering the input cosmological parameters of the simulation}

We test the model defined in Section~\ref{sec:model}
on the mocks described in Section~\ref{sec:mocks}.
In our setup, no noise has been applied to the CMB maps. However, as explained above, we are use a single box for the galaxies, so we are subject to a single realization of the galaxy power spectrum and shot noise. Therefore while we expect to be on average $<1 \sigma$ away from the true value in the mocks, residual noise fluctuations mean that perfect agreement should not be expected.  As described below, when we add realistic reconstruction noise to the CMB lensing maps, the recovered parameters are typically $\sim 1 \sigma$ away from their true value, as expected.

We first verify that we recover the correct input cosmology
on noiseless mocks, using the true $W^{{\rm dir},g}(z)$ and $W^{{\rm xc}, g}(\chi)$
of the simulation (as given in Fig.~\ref{fig:dndz_comparison}).
Our goal for validating the pipeline is to recover
$S_8$, $\sigma_8$ and $\Omega_m$ to within 0.5$\sigma$
(as measured from the marginalized one-dimensional posteriors),
although in some cases the biases in $\sigma_8$ and $\Omega_m$
are slightly higher, possibly due to noise in the large-scale
modes of the simulation that most of our constraining power comes from.

In these cosmology runs, we fix the other cosmological
parameters to their true values in the simulation.
This is a conservative test, as the errors in the data
will be larger for two reasons: first, marginalizing
over the uncertain redshift distribution (marginally) increases the errors (Section~\ref{subsec:dndz_marg}) and second,
in our fiducial analysis we shall keep $\Omega_m h^3$ fixed (corresponding to a fixed angle of the CMB peaks, see Section \ref{subsec:params_marg}) rather than $h$,
further increasing the uncertainty on $\Omega_m$ and $\sigma_8$.

Our sampled parameters are the (Eulerian) linear bias $b_{1,E}^{\rm eff}$, $\Omega_m$, $s_\mu$, $\log_{10}(\textrm{Shot Noise})$, and $\ln{(10^{10}A_s)}$.
We summarize the priors on these parameters in Table~\ref{tab:prior}.
The priors on $\Omega_m$ and $\log(10^{10} A_s)$ and $b_{1,E}^{\rm eff}$ are chosen to be uniform
and much larger than the constraints
provided by the data; hence the arbitrary choice of where to cut off the uniform priors doesn't matter.
For shot noise, the Gaussian prior on log shot noise
corresponds to 50\% variation from the Poisson value.
Finally, for the magnification parameter, $s_\mu$, we choose a 10\% uncertainty around the measured value from \cite{Krolewski20}.
This is driven by two factors: first, since the WISE depth varies with ecliptic latitude, the measured slope of the galaxy
magnitude distribution at the faint end also varies
by about 5\% (see Fig.~20 in ref.~\cite{Krolewski20});
and second, to remove potential modelling biases
from the fact that the magnification term
maps to larger $k$ where our model
is likely to perform worse.

We sample from the posterior
using  the MCMC sampler developed for \texttt{CosmoMC} \cite{Lewis:2002ah,Lewis:2013hha} and tailored for parameter spaces with a speed hierarchy,
as implemented in the \texttt{Cobaya}\footnote{\url{https://cobaya.readthedocs.io}} package \cite{Cobaya,CobayaSoftware}.
We determine chain convergence using a generalized
version of the $R-1$ Gelman-Rubin statistic \cite{GelmanRubin92,Lewis:2013hha},
determining chains to have converged once $R-1 < 0.1$.
We remove the first 30\% of the chains from all analyses
as burn-in.

\begin{table}[]
\centering
\begin{tabular}{c|c}
Parameter & Prior \\
\hline
$\Omega_m$ & $[0.1, 0.9]$ \\
$\ln(10^{10} A_s)$ & $[1.0, 4.0]$ \\
$\log_{10}({\textrm{Shot Noise}})$ & $\mathcal{N}$($\log_{10}(\bar{n}^{-1}), 0.2^2$) \\
$b_{1,E}^{\rm eff}$ & $[-10, 10]$ \\
$s_\mu^{\rm blue}$ & $\mathcal{N}(0.455, 0.0455^2)$ \\
$s_\mu^{\rm green}$ & $\mathcal{N}(0.653, 0.0653^2)$ \\
$s_\mu^{\rm red}$ &  $\mathcal{N}(0.842, 0.0842^2)$ \\
\end{tabular}
\caption{The priors on the model parameters, discussed in the text.  The priors on the cosmological
parameters and bias are uniform and $\mathcal{N}(x, \sigma^2)$ denotes a Gaussian distribution with mean $x$ and standard deviation $\sigma$.  The Gaussian prior on log shot noise
allows 50\% variation from the Poisson value.}
\label{tab:prior}
\end{table}

Using a covariance matrix appropriate for the data with $f_{\rm sky} = 0.586$ we find that we recover the correct input cosmology with $\ell_{\rm max} = 300$ for the blue and green samples, while for the red sample, we choose the more conservative $\ell_{\rm max} = 250$.
For all samples, we use $\ell_{\rm min, auto} = 100$ and $\ell_{\rm min, cross} = 20$, exactly
the same scale cuts that we use on data.

\subsection{Marginalizing over the uncertain redshift distribution}
\label{subsec:dndz_marg}

Due to the noisy nature of redshift distribution measurement $\hat{W}^{{\rm xc}, g}(\chi)$, it is important to propagate the uncertainty on redshift distribution, to the uncertainty in cosmological parameters\footnote{In principle we could do this for the cross-correlation redshifts by including all of the measured cross-spectra in the inference, along with a model for $dN/dz$ and the spectroscopic populations.  This would significantly increase the complexity of the model, and would not address the direct-match $dN/dz$ estimates, so we leave this for future work.}.  We shall do this by a simple averaging procedure as described below.

First, we generate $W^{{\rm xc}, g}(\chi) \propto b_1(z) dN/d\chi$ samples consistent with the measurement noise using B-splines following Section 5.2 in ref.~\cite{Krolewski20}. These samples are obtained by considering the measured $\hat{W}^{{\rm xc}, g}(\chi)$ and its noise covariance. A Gaussian random realization with the correct noise covariance is generated many times (one for each sample), then this noise realization is added to the measured $\hat{W}^{{\rm xc}, g}(\chi)$ and finally a smooth B-spline with positivity constraint and curvature penalty is fit to the resulting $W^{{\rm xc}, g}(\chi)$ sample. This procedure generates (bias-weighted) redshift distributions that are consistent with the data and whose density in the set of possible distributions is proportional to their probability of being the correct one given the data\footnote{To be precise, given the cross-correlation between the unWISE galaxies and the spectroscopic data used to determine $\hat{W}^{{\rm xc}, g}(\chi)$. Some amount of information about the redshift distribution is also contained in the goodness of fit to the $C_\ell^{gg}$ and $C_\ell^{\kappa g}$ data, and this justifies the further weighting by the posterior value outlined in the next paragraph.}.

Next, we run a number of MCMC chains (each sampling over the 5 parameters listed in the previous section), one for each sample ${W}^{{\rm xc}, g}(\chi)$ rather than just the best-fit $\hat{W}^{{\rm xc}, g}(\chi)$.
Finally we combine the chains of each $W^{{\rm xc}, g}(\chi)$
sample, weighting each by the posterior value at the maximum (MAP)
for each sample $P(\rm{model\, at\, MAP} | \rm{data})$.

We use 20 samples for the averaging, and have found that this is sufficient to propagate the effect of redshift uncertainty, because of the considerable overlap of the contours with different samples.
As usual, we test this procedure
on simulations before applying it to
data. We generate $W^{{\rm xc}, g}(\chi)$ samples
for the mocks by adopting a similar error covariance matrix to what we found in the data 
and then sampling from $W^{{\rm xc}, g}(\chi)$. We further apply a noise bias correction as described in the next subsection, while noting that alternative methods such as those described in \cite{Myles:2020dyq, Hadzhiyska:2020xob} can be used instead.

In Fig.\ \ref{fig:mock_grn_contours}, we show the parameter posteriors before and after redshift uncertainty marginalization for the green sample in the CrowCanyon2 simulation (see Fig.\ \ref{fig:mock_two_contours} for equivalent plots for the blue and red samples).

We find that $W^{{\rm xc}, g}(\chi)$ marginalization has a small impact on the
marginalized $\Omega_m$ and $\sigma_8$ constraints, changing them by $< 15\%$,
while it does cause a $\sim 20-50\%$ increase in the $S_8$ errors (with a larger impact on the blue sample than green or red). More generally, we find that $W^{{\rm xc}, g}(\chi)$ marginalization affects nuisance parameters such as $b_{1,E}^{\rm eff}$ more than our cosmological parameters of interest. 

Some previous analyses have adopted a more compact parametrization of the redshift uncertainty, for example by marginalizing over a shift and a width of the redshift distribution.  In our specific setup, we find that such a marginalization is unable to properly account for the redshift errors. For example, when analyzing the mocks with one of the $W^{{\rm xc}, g}(\chi)$ samples rather than the true one, the addition of shift and width parameters don't appear to decrease the size of the bias. For this reason, we adopt the previous setup.

\subsection{Noise bias correction}
\label{sec:noise_bias}

There is one further subtlety when computing a theoretical prediction for the $C_\ell$ by using a noisy $\hat{W}^{{\rm xc}, g}$, where we have further imposed a positivity constraint. Because the theoretical $C_\ell$ are a non-linear function of the value of the $dN/dz$ at a particular redshift, a noise bias is introduced\footnote{If we allowed a negative $dN/dz$ and fully marginalized over the $dN/dz$ noise realizations, there would be no need for a noise bias correction. However, given our smoothness prior and the positivity constraint, we correct for the small bias introduced with the method described in this session. Note that most ``low-dimensional'' $dN/dz$ marginalizations, such as those using a shift and a width parameter don't fully marginalize over possible noise realizations, and therefore introduce a noise bias that should in principle be corrected for.} when computing the $C_\ell$ given a $W^{{\rm xc}, g}(\chi)$. This is easy to understand intuitively, for example in the case of the auto-correlation $C_\ell^{gg}$: suppose that true $dN/dz$ was zero in a particular redshift range. However, because of noise in its measurement and the positivity constraint, the measured $\hat{W}^{{\rm xc}, g}(\chi)$ will either positive (due to a positive noise fluctuation) or zero. Therefore, a region with no galaxies will always appear to have a non-negative noise floor in terms of effective number of galaxies, thus biasing the theoretical $C_\ell^{gg}$. This is quite general, and a noise bias is usually introduced when taking a non-linear function of a noisy quantity.

We can correct for this by Monte-Carlo, by computing the $C_\ell$ with 1000 noise realizations in $W^{{\rm xc}, g}(\chi)$ and comparing the result to the $C_\ell$ computed at the fiducial $\hat{W}^{{\rm xc}, g}(\chi)$.
More explicitly: for each of the galaxy auto and cross-correlation (separately), and for each tomographic bin we compute the quantity 
\begin{equation}
    \left( \Delta C_\ell \right)_\text{noise\ bias} = \left \langle C_\ell^\text{theory} \right \rangle_{W^{{\rm xc}, g}(\chi) {\rm \ samples}} - \left( C_\ell^\text{theory} \right)_{{\rm fiducial\ } \hat{W}^{{\rm xc}, g}(\chi)} 
\end{equation}
Where the ``fiducial'' $\hat{W}^{{\rm xc}, g}(\chi)$ refers either to the best-fit measured one on the data\footnote{After smoothing with a B-spline.}, or the actual $W^{{\rm xc}, g}(\chi)$ in the mocks (they are very similar, but we treat each case separately and self-consistently). Similarly, we have created a set of sample $W^{{\rm xc}, g}(\chi)$ for each of the mocks and data, and for each tomographic bin.
Before cosmological inference, we correct the theory $C_\ell$ of both the mocks and the real data by subtracting the noise bias:
\begin{equation}
   \left( C_\ell \right)_\text{corrected} = C_\ell^\text{theory} - \left( \Delta C_\ell \right)_\text{noise\ bias}
\end{equation}

We find that this noise bias correction is significant for correctly interpreting $C_\ell^{gg}$, while it's negligible for $C_\ell^{\kappa g}$. For completeness we always include this correction, except when performing tests on mocks with $W^{{\rm xc}, g}(\chi)$ fixed to its fiducial (true) value.
As validation, we note that the median
of the posteriors on mocks is nearly unchanged 
between the case with the true $dN/dz$
and the samples of $dN/dz$.
While the noise bias correction is slightly
different between the mocks and the data (and we always use the appropriate one in any analysis),
if we use the mock-based noise bias instead of the data-based
one when constraining parameters with the data, we find shifts of $<0.5\sigma$.

\begin{table}[]
\small
\centering
\begin{tabular}{c|cc|cc|cc}
Test & $\Omega_m$  & Bias/$\sigma$
& $\sigma_8$  & Bias/$\sigma$
& $S_8$  & Bias/$\sigma$ \\
\hline
True value   & $0.3092$ & -- & $0.822$ & -- & $0.835$ & -- \\
Fix geom. & $0.3196 \pm 0.016$ & 0.63 & $0.826 \pm 0.025$ & -0.17 & $0.852\pm 0.016$ & 1.08\\
Fix geom., fix $\Omega_m h^3$& $0.3305 \pm 0.032$ & 0.67 & $0.806 \pm 0.046$ & -0.34 & $0.847\pm 0.017$ & 0.7\\
Fix geom., sample $dN/dz$& $0.3192 \pm 0.016$ & 0.61 & $0.820 \pm 0.031$ & -0.07 & $0.846\pm 0.028$ & 0.43\\
\rowcolor{gray!20}
Fix $\Omega_m h^3$, sample $dN/dz$ & $0.3244 \pm 0.030$ & 0.51 & $0.812 \pm 0.046$ & -0.22 & $0.844\pm 0.026$ & 0.37\\
\end{tabular}
\caption{Performance of the model on mock data for the blue sample.  The tests are described in the text.  The test in the final row is also done with fixed geometry, which we omit from the row heading for compactness.  For each test we give the median and uncertainty of each recovered parameter, plus the offset of the median from the true value as a fraction of the uncertainty (i.e.\ the bias in units of $\sigma$).  The grey row indicates the assumptions for our fiducial constraints. Values are quoted as the median and $1/4$ of the difference between the $2.5^{\rm th}$ and $97.5^{\rm th}$ percentiles. }
\label{tab:blue_test}
\end{table}

\begin{table}[]
\small
\centering
\begin{tabular}{c|cc|cc|cc}
Test & $\Omega_m$  & Bias/$\sigma$
& $\sigma_8$  & Bias/$\sigma$
& $S_8$ & Bias/$\sigma$ \\
\hline
True value & $0.3092$ & -- & $0.822$ & -- & $0.835$ & -- \\
Free geom. & $0.311 \pm 0.013$ & 0.18 & $0.833 \pm 0.017$ & 0.62 & $0.848 \pm 0.016$ & 0.86  \\
Fix geom. & $0.3111 \pm 0.011$ & 0.17 & $0.832 \pm 0.020$ & 0.49 & $0.847  \pm 0.014$ & 0.9\\
Fix geom., fix $\Omega_m h^3$ & $0.3178 \pm 0.020$ & 0.43 & $0.820 \pm 0.034$ & -0.07 & $0.844  \pm 0.015$ & 0.62  \\
Fix geom., sample $dN/dz$ & $0.3102 \pm 0.012$ & 0.09 & $0.831 \pm 0.023$ & 0.39 & $0.845\pm 0.018$ & 0.58 \\
\rowcolor{gray!20}
Fix $\Omega_m h^3$, sample $dN/dz$ & $0.3167 \pm 0.020$ & 0.37 & $0.820 \pm 0.033$ & -0.07 & $0.843  \pm 0.017$ & 0.47 \\
\end{tabular}
\caption{As for Table \ref{tab:blue_test}, but for the mock green sample.}
\label{tab:green_test}
\end{table}

\begin{table}[]
\small
\centering
\begin{tabular}{c|cc|cc|cc}
Test & $\Omega_m$  & Bias/$\sigma$
& $\sigma_8$  & Bias/$\sigma$
& $S_8$  & Bias/$\sigma$ \\
\hline
True value & $0.3092$ & -- & $0.822$ & -- & $0.835$ & -- \\
Fix geom.  & $0.3031 \pm 0.017$ & -0.37 & $0.869 \pm 0.047$ & 0.99 & $0.875  \pm 0.043$ & 0.94\\
Fix geom., fix $\Omega_m h^3$ & $0.2975 \pm 0.032$ & -0.37 & $0.877 \pm 0.061$ & 0.9 & $0.877  \pm 0.045$ & 0.94  \\
Fix geom., sample $dN/dz$ & $0.3012 \pm 0.018$ & -0.45 & $0.871 \pm 0.051$ & 0.96 & $0.873\pm0.049$ & 0.77 \\
\rowcolor{gray!20}
Fix $\Omega_m h^3$, sample $dN/dz$ & $0.2983 \pm 0.033$ & -0.33 & $0.875 \pm 0.064$ & 0.83 & $0.874  \pm 0.047$ & 0.83 \\
\end{tabular}
\caption{As for Table \ref{tab:blue_test}, but for the mock red sample.}
\label{tab:red_test}
\end{table}

\subsection{Results on mocks}

\begin{figure}
    \centering
    \includegraphics[width=\textwidth]{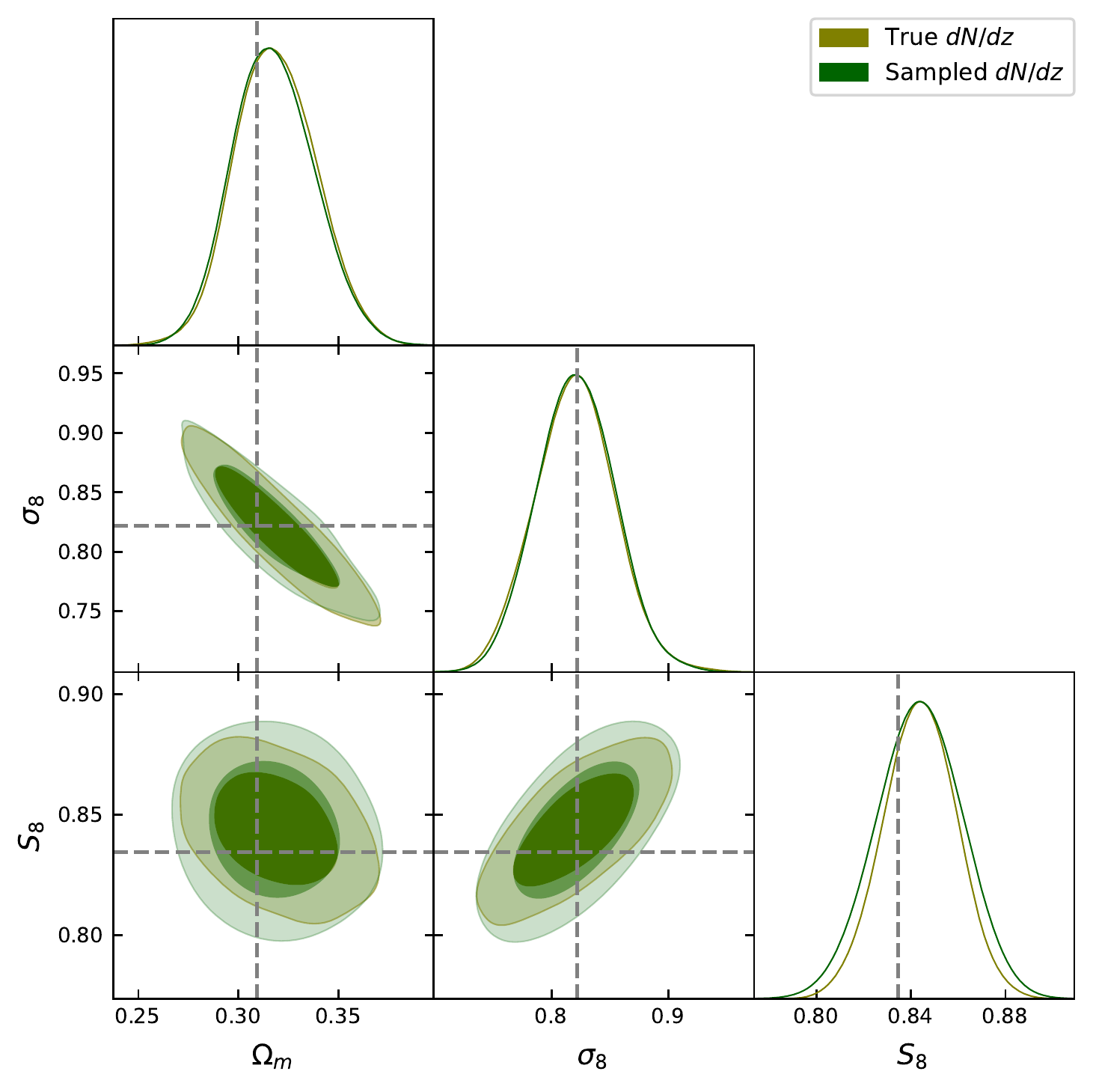}
    \caption{Posteriors for the mock green sample, showing the key cosmological parameters: $\Omega_m$ and $\sigma_8$ plus the derived parameter $S_8$.  We show the results for a fixed $dN/dz$ and for sampling over the $dN/dz$ uncertainty, as described in the text.  Dashed grey lines show the ``true'' values of the parameters in the simulation, though we expect some scatter due to sample variance in the initial conditions. }
    \label{fig:mock_grn_contours}
\end{figure}

The parameter constraints from mocks for the green sample, using the true $W^{{\rm dir},g}(z)$ and $W^{{\rm xc}, g}(\chi)$ of the simulation, are shown in Fig.~\ref{fig:mock_grn_contours}
(similar results for the blue and red samples can be found in Fig.~\ref{fig:mock_two_contours}).
The key result from this test
is the discrepancy between
the true cosmological parameters
and the median of the recovered
values, which we refer to as the bias in the parameters.
Our fiducial estimate
for the recovered parameters
is the median rather than the maximum a posteriori (MAP), although we find that the two are very similar.

We summarize the one-dimensional posteriors
on $\Omega_m$, $\sigma_8$ and $S_8$ for different scenarios in Tables~\ref{tab:blue_test}, \ref{tab:green_test} and \ref{tab:red_test} for the blue, green and red samples respectively.
Recall this test uses noiseless CMB lensing maps, though we use only a single realization of the N-body simulation so we expect some deviation from the input cosmology. However we expect the bias to be $<1 \sigma$ since the overall noise in our simulations is less than that of real data (the CMB lensing noise is a large part of our error budget). We find that the parameters of interest are recovered to better than $0.5\sigma$ for the blue and green samples in our fiducial setup.  For the red sample, the modeling is more challenging, and we find possible biases in $S_8$ up to $0.8\sigma$. The red sample doesn't provide
much constraining power, so we shall not explore this further. In the combined analysis, we add a systematic error correction to the red sample to account for the larger modeling error.  Specifically, for each bin in the auto and cross-correlation, we add in quadrature a systematic error equal to 0.94 times the statistical error, as derived from the third row of Table~\ref{tab:red_test}.  This conservatively assumes that the entirety of the discrepancy in Table~\ref{tab:red_test} arises from modeling errors, although some of it may come from cosmic variance on the single simulated sky that we used.

We also test the cosmology pipeline when Planck-appropriate
noise is added to the $\kappa$ map.
Unsurprisingly, here we find $\sim 1 \sigma$ biases,
since a single noise realization in the data should cause scatter by $\sim 1\sigma$ in the parameters. However, this test
does allow us to verify that the model
provides a good fit to the simulation data, with $\chi^2$/d.o.f. $\sim$ 1.

Additionally, we find that the maximum a posteriori
cosmological parameters are quite similar to the medians
of the marginalized posterior quoted in Tables~\ref{tab:blue_test}-\ref{tab:red_test}.
Thus we opt not to use the MAP and corresponding
MAP-based confidence intervals as advocated
by \cite{Joachimi21} for weak lensing constraints,
where the likelihood surface is more non-Gaussian.

We additionally try a number of variations
on this model and find they do not 
affect our results.
We try allowing the higher-bias
power spectra to vary with cosmology;
allowing the scaling factor on $b_{\rm 2,L}$
to vary or allowing for a linear offset in $b_{\rm 2,L}(z)$; measuring $b_{\rm sml}$ in on larger
(more linear) scales ($>30\,h^{-1}$Mpc); using more points
in $z$ to interpolate $b_{\rm sml}(z)$;
and using different simulation boxes
with different seeds.

\section{Cosmological constraints}
\label{sec:cosmology}

\subsection{Blinding strategy}

Blinding is important to reduce the chances of introducing spurious biases (such as ``confirmation bias'') through our analysis choices, choice of parameters, scale cuts, data selection etc. With this in mind, we adopt the following strategy: we don't blind the bandpowers of the auto and cross-correlation since those were already published \cite{Krolewski20}. In that work, however, we did not attempt any cosmological inference.  A large number of null and systematic tests were performed to confirm the robustness of the bandpowers.
Next, for each tomographic bin we have created mocks based on simulations to match the redshift distribution and bias evolution of the sample, as described in Section \ref{sec:mocks}.  Then we tested the model described in Section \ref{sec:model}, and we chose scale cuts and which parameters to vary based solely on the mocks.  After all of the null tests were performed and passed, and the analysis choices fixed based on getting unbiased cosmology from the mocks (i.e. recovering the input parameters of the underlying dark matter simulation), we have decided to ``unblind'' the data, by running MCMC chains on the measured bandpowers with the same setup used on the mocks. We have not modified the pipeline or analysis choices after unblinding.
In summary, throughout the analysis, we have not been blinded to the bandpowers of the auto and cross-correlations, but we have been fully blinded to the cosmology.

\subsection{Systematics checks}

In Section 7 of the companion paper \cite{Krolewski20}, we performed a large number of null and systematics tests to validate the robustness of our measurement. These tests are summarized in Fig. 12 of ref.~\cite{Krolewski20} and here we briefly comment on them.

First we have tested the isotropy of the signal: no significant variation was observed when restricting the analysis to the BOSS footprint, which is the one used to measure the cross-correlation redshifts. Since the unWISE imaging depth is spatially dependent\footnote{However our magnitude cuts ensure that completeness is close to 100\% everywhere on the footprint.}, this is an important test. Next we studied the impact of the range of scales involved in determining $\hat{W}^{{\rm xc}, g}(\chi)$ from the cross-correlation with BOSS and eBOSS, again finding negligible changes. Similarly, the sky is split into two halves at Galactic longitude = 155$^\circ$ and both the power spectra and clustering redshifts are consistent between the two halves. The use of systematic weights or changing in the magnification bias slope $s_\mu$ have likewise a very small impact on the amplitude of the auto and cross-correlations. 
We have also verified that more restrictive scale cuts have negligible impact. Additionally, extra masking around stars and the presence of the transfer function described in Section \ref{sec:data} don't impact our results.
Finally, in Section 7.2 of ref.~\cite{Krolewski20}, we argued that our measurement should be robust to possible foregrounds in the CMB lensing map that are correlated with the unWISE sample, based on the estimates of \cite{vanEngelen:2013rla,Schaan:2018tup, Sailer:2020lal}, together with the null tests in \cite{PlanckLens18}, Section 4.5. As a further test we also repeated the cross-correlation with CMB lensing maps obtained from tSZ-deprojected temperature maps, getting consistent results. 

In summary, in all of the tests that we have performed, the statistical error has always been  dominant over the shift in the auto and cross-correlation amplitude, giving confidence in the robustness of the $C_\ell^{\kappa g}$ and $C_\ell^{gg}$ used in this analysis.

\subsection{Parameter marginalization}
\label{subsec:params_marg}

Our primary focus is determining the low-redshift values of the cosmological parameters that set the amplitude of lensing. Since our measurements are only at low redshift, we are unable to determine the full set of cosmological parameters as measured, for example, by the primary CMB. First, our measurements are insensitive to (and independent of) the value of the optical depth to reionization $\tau$. Therefore we will set it to the fiducial value whenever calculating a power spectrum with $\texttt{CAMB}$ and drop it from our parameter set, noting that the results won't be affected by the particular value chosen. Second, we note that since we are using projected quantities\footnote{And hence the scale and size of the baryon acoustic oscillations (BAO) have been erased.} at low redshift, we are only sensitive to the total low-redshift matter density $\Omega_m$ and not separately to its components of cold Dark Matter ($\Omega_c$) and baryons ($\Omega_b$). Therefore we will use only one parameter $\Omega_m = \Omega_c + \Omega_b + \Omega_{\nu}$, and fix\footnote{Note that, on the scales of interest, we are sensitive to baryons primarily through their gravitational influence, which is degenerate with the effect of the Dark Matter. When running \texttt{CAMB} we choose to set the value above, but point out that our results are largely independent of this choice within the range of currently accepted values.} $\omega_b \equiv \Omega_b h^2 = 0.02242$. Third, our measurement of low-redshift amplitude is not yet at the level that can allow a detection of neutrino masses given the current bounds, and therefore we set the neutrino density to its value predicted by the minimum mass, normal hierarchy scenario ($\sum m_\nu = 0.06$ eV). 

Finally, our measurements of the lensing amplitude are quite degenerate with the distance to the unWISE redshift, i.e.\ the Hubble parameter $h$. To break this degeneracy, we will use a measurement of the angular size of the sound horizon at recombination from the CMB, $\theta_\star$. This is one of best-measured and most robust quantities from the primary CMB and its measurement is purely geometrical since it's determined by the angular position of the peaks in the CMB temperature and polarization power spectra. Therefore it's largely independent of the particulars of the cosmological model and the detailed physics of the CMB. The Planck satellite provides a remarkable 0.03\% measurement of $\theta_\star = 1.04109 \pm 0.00030$ \cite{2018arXiv180706209P}.
A quick calculation shows that within the $\Lambda$CDM model, $\theta_\star$ is primarily determined by the product $\Omega_m h^3$ \cite{Percival:2002gq}, leading to a geometrical 0.3\% constraint on $\Omega_m h^3 = 0.09633 \pm 0.00029$ \cite{2018arXiv180706209P}. It is important to note that this measurement is largely independent of the complex physics that determines the CMB power spectrum as well as any possible observational systematics affecting the broad-band CMB power spectrum. Therefore, even within our philosophy of being as independent as possible from the physics of the early Universe and of the CMB, we have decided to fix $\Omega_m h^3$ in our fiducial analysis. 

The only residual non-trivial dependence on Planck data and the physics of the CMB is through the scalar spectral index $n_s$. While in the current analysis we fix it to the best-fit Planck value ($n_s =  0.9665 \pm 0.0038$, determined to 0.4\% and therefore with negligible impact on our uncertainty estimation), future measurements will either allow $n_s$ to be measured internally, or alternatively, low-redshift measurements of $n_s$ such as from the galaxy power spectrum can be used instead.

\subsection{Results}
\label{sec:results}

\begin{figure}
    \centering
    \resizebox{\columnwidth}{!}{\includegraphics{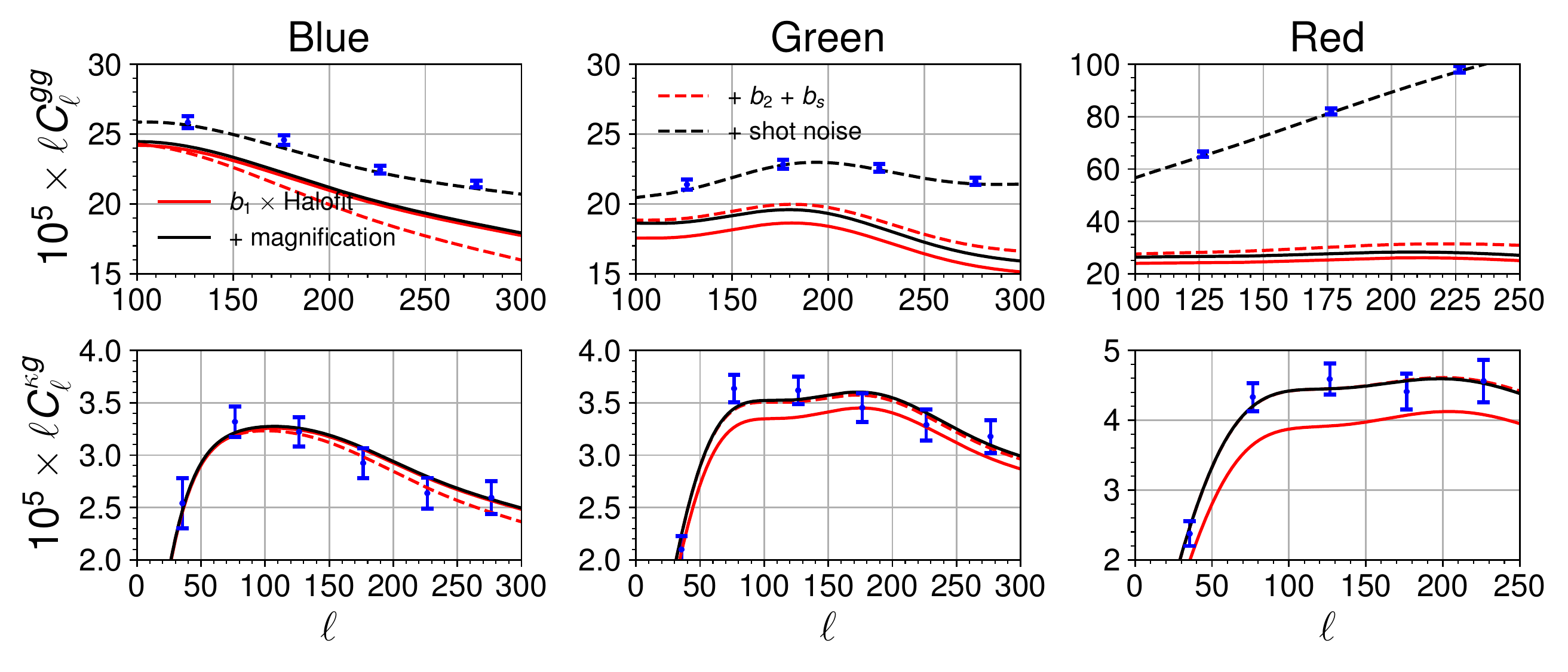}}
    \caption{Comparison between the data and model with MAP parameters
    from the individual fits
    with fixed $\Omega_m h^3$.
    The power spectra are separated
    into various components: the dominant contribution from $b_1$ times Halofit; the magnification contribution; the contribution from higher bias terms, and for the auto-spectra, shot noise.  The model including correction for noise bias from the redshift distribution (Sec.~\ref{sec:noise_bias}) is negligibly different from the model without the correction.
    }
    \label{fig:data_vs_model}
\end{figure}

\begin{figure}
    \centering
    \resizebox{\columnwidth}{!}{\includegraphics{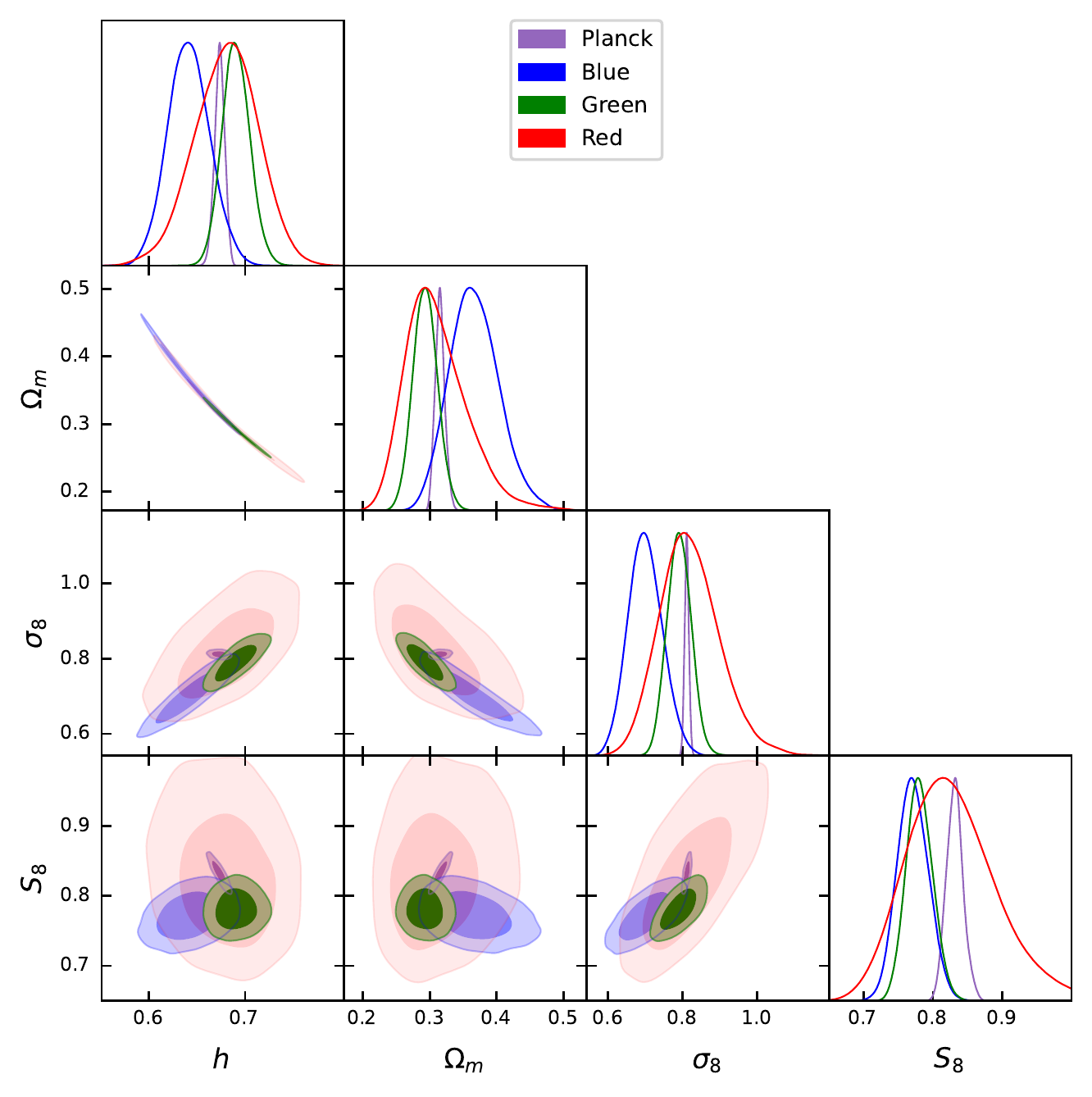}}
    \caption{Constraints on the cosmological parameters from the unWISE blue, green and red samples, compared to those from Planck (temperature, polarization and lensing; purple lines and contours).  In all cases the unWISE contours include marginalization over $dN/dz$ uncertainty.  The tight correlation between $\Omega_m$ and $h$ arises because we fix $\Omega_mh^3$ in our analysis (Section \ref{subsec:params_marg}), thus $\Omega_m$ and $h$ should not be regarded as independent.  }
    \label{fig:data_corner_plot}
\end{figure}

Here we apply the methods discussed in the previous section to the band-powers for the auto and cross correlation measured earlier \cite{Krolewski20} to constrain $\sigma_8$, $\Omega_m$ and $S_8 = \sigma_8 (\Omega_m / 0.3)^{0.5}$, as well as the nuisance parameters for each sample (linear bias, shot noise and magnification slope $s_\mu$). As explained in Section \ref{subsec:params_marg}, in our fiducial analysis we also vary $h$ in such a way as to keep the quantity $\Omega_m h^3$ fixed to its best-fit Planck value, $\Omega_m h^3 = 0.09633$, as inferred from the angular position of the sound horizon at recombination $\theta_\star$. Therefore, we are also able to derive a posterior on $h$.

\begin{table}[]
\small
\centering
\begin{tabular}{c|c|c|c}
Sample & $\Omega_m$ 
& $\sigma_8$ 
& $S_8$ \\
\hline
Blue, fix $h$ & $0.345 \pm 0.020$ & $0.734 \pm 0.029$ & $0.785 \pm 0.016$ \\
Blue, fix $\Omega_m h^3$ & $0.370 \pm 0.036$ & $0.699 \pm 0.046$ & $0.776 \pm 0.018$ \\
Blue, sample $dN/dz$ & $0.342 \pm 0.019$ & $0.728 \pm 0.029$ & $0.777\pm 0.020$ \\
\rowcolor{gray!20}
Blue, fix $\Omega_m h^3$, sample $dN/dz$ & $0.368 \pm 0.037$ & $0.697 \pm 0.045$ & $0.771 \pm 0.022$ \\ \hline
Green, fix $h$ & $0.298 \pm 0.01$ & $0.780 \pm 0.021$ & $0.776 \pm 0.015$ \\
Green, vary $b_2^{\rm off}$ & $0.298 \pm 0.01$ & $0.789 \pm 0.030$ & $0.785 \pm 0.030$ \\
Green, fix $\Omega_m h^3$ & $0.291 \pm 0.018$ & $0.794 \pm 0.032$ & $0.782 \pm 0.014$ \\
Green, sample $dN/dz$ & $0.299 \pm 0.011 $& $0.778 \pm 0.021$ & $0.777 \pm 0.018$ \\
\rowcolor{gray!20}
Green, fix $\Omega_m h^3$, sample $dN/dz$ & $0.293 \pm 0.018$ & $0.792 \pm 0.032$ & $0.781 \pm 0.019$ \\ \hline
Red, fix $h$ & $0.301 \pm 0.019$ & $0.818 \pm 0.062$ & $0.822 \pm 0.061$ \\
Red, fix $\Omega_m h^3$  & $0.305 \pm 0.041$ & $0.816 \pm 0.072$ & $0.822 \pm 0.062$ \\
Red, sample $dN/dz$ & $0.301 \pm 0.023$ & $0.814 \pm 0.064$ & $0.816 \pm 0.064$  \\
\rowcolor{gray!20}
Red, fix $\Omega_m h^3$, sample $dN/dz$ & $0.305 \pm 0.046$ & $0.813 \pm 0.080$ & $0.822 \pm 0.066$  \\ \hline
Blue + green, fix $\Omega_m h^3$ & $0.302 \pm 0.018$ & $0.781 \pm 0.032$ & $0.785 \pm 0.013$  \\ 
\rowcolor{gray!20}
Blue + green, fix $\Omega_m h^3$, sample $dN/dz$ & $0.307 \pm 0.018$ & $0.773 \pm 0.029$ & $0.782 \pm 0.015$  \\ 
+ BAO & $0.307 \pm 0.007$ & $0.772 \pm 0.018$ & $0.781 \pm 0.015$  \\
\hline
Blue + green + red, fix $\Omega_m h^3$ & $0.309 \pm 0.017$ & $0.774 \pm 0.027$ & $0.785 \pm 0.013$  \\ 
\rowcolor{gray!20}
Blue + green + red, fix $\Omega_m h^3$, sample $dN/dz$ & $0.307 \pm 0.018$ & $0.775 \pm 0.029$ & $0.784 \pm 0.015$  \\ 
+ BAO & $0.307 \pm 0.007$ & $0.775 \pm 0.018$ & $0.784 \pm 0.016$  \\
\hline
\end{tabular}
\caption{Parameter constraints for the unWISE-Planck cross-correlations.  The different variations on the analysis are discussed in the text, with our primary constraints being for fixed $\Omega_m\,h^3$ including the sampling over the $dN/dz$ uncertainty.  We caution that constraints on $\Omega_m$ and $\sigma_8$ are correlated (see Fig.~\ref{fig:data_corner_plot}), and this degeneracy needs to be broken by the addition of external data (e.g.\ SDSS I, II and III BAO measurements \cite{Ross15,Alam21} plus 6dF BAO \cite{Beutler12} in the last row).
}
\label{tab:data}
\end{table}

Our main results are summarized in Table \ref{tab:data} and shown in Fig. \ref{fig:data_corner_plot} (the posteriors for the full parameter set and our fiducial sample can be found in Appendix \ref{app:posteriors}).  This figure makes it clear that we constrain the combination $S_8=\sigma_8(\Omega_m/0.3)^{0.5}$ more robustly than either $\Omega_m$ or $\sigma_8$ alone, with those parameters being only loosely constrained in the degeneracy direction. As is also apparent from Fig. \ref{fig:data_corner_plot}, the green sample is the most constraining: this is a combination of it being the highest $S/N$ bin in the cross-correlation with Planck lensing and it being at high enough redshifts that non-linearities at a fixed angular scale are less important\footnote{Due both to the increase of $k_{\rm NL}$ with redshift, and the fact that a fixed physical scale, $k$, projects to a larger $\ell$ at higher redshift [$\ell \sim k \chi(z)$].} and therefore it can be modeled to higher $\ell_{\rm max}$.  
The blue sample is the most affected by non-linearities, rendering the modeling more challenging, while the large shot noise and the importance of non-linear biases of the red limit its constraining power. Moreover, the green sample is the most well-characterized in terms of both redshift distribution and sample properties.  Broadly speaking, we find that the three samples are statistically consistent, even though some of the degenerate parameters for the green and blue are slightly offset. Moreover, as we will see, the parameter determining the amplitude of lensing, $S_8$, is very similar between all of them.
Given that the systematics, redshift, density, bias and amount of non-linearity are very different for our three samples, this is a very important test and highlights the robustness of our measurement. 

\begin{figure}
    \centering
    \resizebox{\columnwidth}{!}{\includegraphics{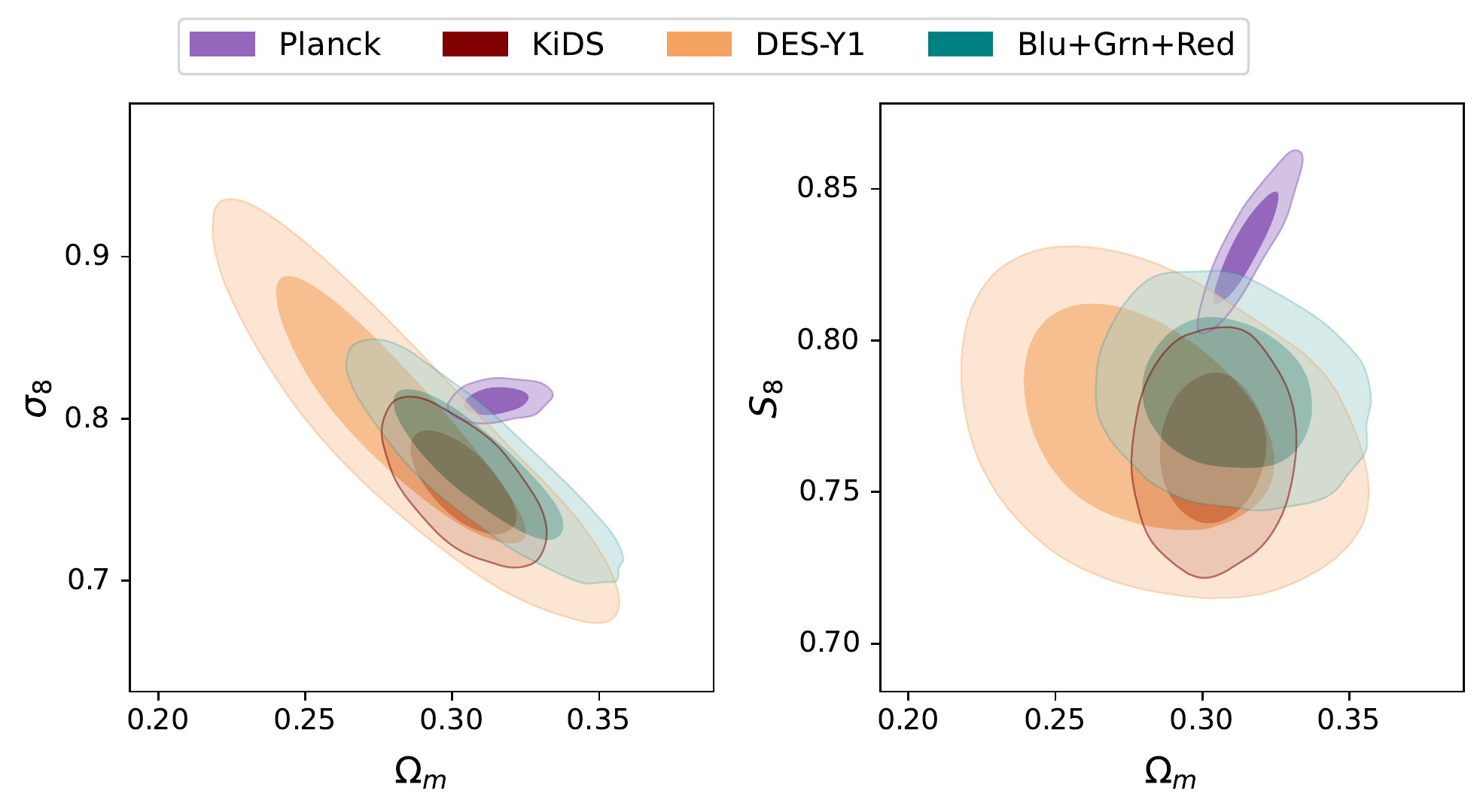}}
    \caption{Constraints on the matter density and power spectrum amplitude from the unWISE blue and green samples, compared to those from Planck (purple \cite{PlanckParams18}), KiDS+BOSS+2dFlens (maroon \cite{kids1000}) and DES (light brown \cite{Abbott:2017wau}).  We have importance weighted the KiDS and DES chains to impose a similar prior on $\Omega_m h^3$ as used for unWISE, though that has only a minor effect on the contours.  If we add SDSS I-III and 6dF BAO data to our Blu+Grn constraints the constraints shrink dramatically in the $\Omega_m$ direction, tigthening around $\Omega_m\approx 0.3$, however the constraint on $S_8$ is unchanged.
    }
    \label{fig:data_sml_corner_plot}
\end{figure}

The best-constrained, `lensing' parameter $S_8$ (roughly corresponding to the combination perpendicular to the degeneracy direction in the $\Omega_m - \sigma_8$ plane) appears to be below the values from the Planck primary CMB measurements. In fact, Planck reports $S_8= 0.832 \pm 0.013$ (including CMB lensing) \cite{PlanckParams18}. For the green sample, we find $S_8 = 0.781 \pm 0.019$, showing a ``tension'' with Planck to $\approx 2.2 \sigma$ when adding the statistical uncertainties in quadrature.  For the blue sample, we find a very similar results of $S_8 = 0.771 \pm 0.022$, or a tension with Planck of about $2.4\sigma$. For the red sample we find a consistent value of $S_8 = 0.822 \pm 0.066$; the uncertainty is considerably larger than the other two samples due to the inclusion of modelling systematic error and the considerably lower number density.  The model with maximum a posteriori parameters is compared to the data in Fig.~\ref{fig:data_vs_model}.  As described earlier we do not use the Planck $\kappa$ auto-correlation in our fits.  The auto-correlation alone gives $S_8 = 0.818\pm 0.018$, once BAO and weak priors are included \cite{PlanckLens18}.  The combined result is $1.5\,\sigma$ lower than this, with a comparable uncertainty.

In our fiducial model, the higher-bias
evolution is fixed using relations found from
simulated dark matter halos \cite{Abidi18}. We could
instead allow the higher-bias evolution to vary
by adding a free constant $b_2^{\rm off}$ to $b_2(z)$ as given by Equation~\ref{eqn:bias_evolution} and Fig.\ 8 of \cite{Abidi18} (with an unconstraining prior on $b_2^{\rm off}$). Adding
this freedom to the green sample increases the errorbar on $S_8$ by a factor of 2 and decreases the tension with Planck to $1.17\sigma$ (using the upper error bar of $0.038$ rather than the symmetric errorbar of $0.030$ reported in Table~\ref{tab:data}, to account for the asymmetry in the marginalized $S_8$ posterior).  This is because
$b_2^{\rm off}$ is quite degenerate with $S_8$, and in particular $b_2^{\rm off} \sim -2$ can restore $S_8$ to nearly its Planck value while still providing a good fit to the data.
However, our data are quite consistent with $b_2^{\rm off} = 0$ ($b_2^{\rm off} = -0.26 \pm 1.02$), and adding
$b_2^{\rm off}$ does not improve the goodness of fit,
nor does it reduce the differences between the true
cosmology and the best-fit cosmology in Tables~\ref{tab:blue_test}--~\ref{tab:red_test}.
Moreover, we expect that adding parameters
will always degrade the constraining power of the data, and we emphasize that our simulation tests
were carried out with $b_2^{\rm off} = 0$ and
do not detect nonzero $b_2^{\rm off}$.
Furthermore, we require a relatively large value
of $b_2^{\rm off} = -2$ to bring our data into concordance with the Planck value of $S_8$.  
Based on tests from simulations, Refs.~\cite{Zennaro21b} and~\cite{Barreira21} suggest
that $b_2^{\rm off} \lesssim 0.5$ for galaxy
samples selected using stellar mass or star formation rate,
though Ref.~\cite{Barreira21} estimate $b_2^{\rm off}$ from fits to data in the literature
and find $b_2^{\rm off} \sim -2$ for the BOSS DR12
measurements of Ref.~\cite{Ivanov20}.

Given the difficulty in breaking the $\sigma_8 - \Omega_m$ degeneracy without external datasets, and the consistency in the value of $S_8$, we have decided to combine the green, blue and red samples and to quote joint constraints  as our fiducial values (we note that the red sample adds hardly any constraining power and the combined result is very similar to the green plus blue result).  We combine the blue, green and red $C_\ell^{gg}$ and $C_\ell^{\kappa g}$ into a single data vector and include a smooth fit to the measured cross-correlation between blue, green and red galaxies (Fig.\ 22 in \cite{Krolewski20}) when computing the joint covariance matrix.  (The blue-green, blue-red and green-red cross-correlations are not included in the data vector.)  The ``blue + green + red'' joint fit has $\chi^2 / \text{d.o.f.}$ = 17.1/16 (i.e.\ a good fit), and we find $S_8 =0.784 \pm 0.015$, which is $\approx 2.4\sigma$ lower than Planck.
We also note that a similar preference for lower $S_8$ than Planck has also been found from cross-correlating optical galaxies with the Planck lensing map \cite{Hang21,Kitanidis21}; with ref.~\cite{Hang21} interpreting their result as a constraint on $\Omega_m^{0.78} \sigma_8 = 0.297 \pm 0.009$.  For the unWISE sample the addition of SDSS I-III and 6dF BAO data \cite{Beutler12,Ross15,Alam21} serves to fix $\Omega_m\approx 0.3$ and breaks the $\Omega_m-\sigma_8$ degeneracy, however it does not alter our constraint on $S_8$ (see Table \ref{tab:data}).  We find that our data are quite consistent with the BAO results, and the best fitting model is a good fit to the joint data.

In general we find good agreement with recent galaxy lensing measurements.  Fig.~\ref{fig:data_sml_corner_plot} compares our measurements to the DES Y1 \cite{Abbott:2017wau} ``$3 \times 2$'' analysis, which includes the auto-correlation of shear and tracer galaxies, as well as their cross-correlation. We find excellent agreement with the DES Y1 results ($S_8 = 0.773^{+0.026}_{-0.020}$) and we are able to further tighten their constraints\footnote{When comparing DES to our results, we should consider that (as explained in the introduction), we have chosen not to include the CMB lensing auto-power spectrum, which would considerably tighten our contours. However, using it would introduce sensitivity to the higher-redshift matter fluctuations and we have decided to avoid it in order to get ``low-redshift''-only constraints. By contrast the galaxy lensing auto correlation is included in the DES results and other ``$3 \times 2$'' analyses.  The Planck lensing autocorrelation is consistent with the primary CMB constraints \cite{PlanckParams18}.}.  We are also fully consistent with the results from the KiDS+BOSS+2dFlens surveys, with $S_8 = 0.766^{+0.020}_{-0.014}$ \cite{kids1000} from a combination of lensing and galaxy clustering.  We also agree with galaxy-galaxy lensing measurements using the BOSS galaxies \cite{Singh20}; our blue+green+red sample has $(\sigma_8/0.8228)^{0.8}(\Omega_m/0.307)^{0.6}=0.95\pm 0.02$ while they find $0.85\pm 0.05(\mathrm{stat})\pm 0.05(\mathrm{sys})$ (their constraint from Planck CMB lensing is consistent with ours but significantly weaker). Similarly ``low'' values of $S_8$ have also been found by the first year cosmic shear analysis of HSC \cite{Hikage:2018qbn}, and by CFHTLenS \cite{Heymans:2013fya}.

While we probe the same physics as galaxy lensing, we do so in a way that is operationally very different. For example, estimation of the lensing potential from CMB maps involves measurements of local statistical anisotropy of a very well-characterized source (the primary CMB), and hence is independent of many issues that could potentially affect shear estimation and its interpretation, such as blending, shear calibration, photometric redshift uncertainty of the sources, etc. The excellent agreement between these measurements is therefore a very important check.

We are unable to attribute the bulk of the $S_8$ tension with Planck to low $\Omega_m$ or low $\sigma_8$, and the situation in the broader literature is also unclear.

The Planck constraint on the matter fraction is $\Omega_m = 0.3166 \pm 0.0084$ when fitting to primary CMB only, and  $\Omega_m = 0.3153 \pm 0.0073$ when including the CMB lensing auto-power spectrum \cite{PlanckParams18}. This can be compared to our tightest measurement from the green sample of $\Omega_m = 0.293 \pm 0.018$, i.e.\ our measurement of $\Omega_m$ is about $1.2\,\sigma$ lower than Planck when combining the error bars. However, while the values of $S_8$ are very comparable, the blue sample has a preference for a higher value of $\Omega_m$ (combined with a lower value of $\sigma_8$) highlighting the difficulty in separating these two parameters along the degeneracy direction.  In fact the best fit to the blue sample at $\Omega_m\approx 0.28$ (the best fit to the green sample) has a $\chi^2/\text{d.o.f.}$ of $6 / 4$, so such a `low' $\Omega_m$ value is an acceptable fit to the blue sample despite the contours in Fig.~\ref{fig:data_corner_plot}.
The combined blue + green + red sample has $\Omega_m = 0.307\pm 0.018$, or about $0.4\, \sigma$ lower than Planck.

Interestingly, there are other hints of a ``low'' value of $\Omega_m$ and $S_8$ in the low-redshift Universe: using a BBN prior on $\omega_b = \Omega_b h^2$ to fix scales in physical units, an analysis of the ``full-shape'' redshift-space BOSS galaxy power spectrum \cite{Ivanov20} reports $\Omega_m = 0.295 \pm 0.010$ and $S_8 = 0.703 \pm 0.045$. In that analysis the low value of $S_8$ is mostly driven by the low value of $\sigma_8 = 0.721 \pm 0.043$, rather than a particularly low $\Omega_m$. Very similar results were obtained with a similar model by refs.~\cite{DAmico20, Colas:2019ret, Troster:2019ean}.

There are also hints that within $\Lambda$CDM the Planck preference for higher $\Omega_m$ is being driven by the high $\ell$ temperature data.  For example when viewed in terms of distances along and across the line of sight (Fig.\ 28 of ref.\ \cite{PlanckParams18}) the PlanckTT+lowE (green points) are shifted to lower $\Omega_m$ by the addition of any (or all) of  CMB lensing, polarization or BAO.  Dropping the $\ell>800$ data but combining with lensing and BAO gives $\Omega_m = 0.3081 \pm 0.0065$ and $\sigma_8 = 0.8058 \pm 0.0063$, both within $1\,\sigma$ of our result. Similarly, WMAP9 results\footnote{Effectively an $\ell < 800$ experiment.} \cite{Hinshaw:2012aka}, find a lower value: $\Omega_m = 0.279 \pm 0.025$ (CMB only; though this moves to $\Omega_m = 0.301 \pm 0.008$ including low-redshift BAO data, leading to a higher $S_8 = 0.8201 \pm 0.025$), together with a slightly higher $h =  0.693 \pm 0.009$. Indeed it appears that the high $\ell$ ($>800$) in the CMB temperature likelihood may be driving the fits to higher values of $\Omega_m$ and lower values of $h$ (see Fig.~81 and Section 3.9.2 of ref.\ \cite{PlanckLik18} for further discussion).
This behaviour of the high-$\ell$ CMB TT is not unique to Planck: a recent joint analysis of WMAP9 and ACT data \cite{Aiola:2020azj} finds a ``high'' value of $\Omega_m = 0.313 \pm 0.016$ and $S_8 = 0.840 \pm 0.030$, even though thanks to the larger error bars, the ``tension'' with our measurement is $< 2\sigma$.  Unfortunately, while suggestive, all of these trends are of low statistical significance. 

\begin{figure}
    \centering
    \resizebox{\columnwidth}{!}{\includegraphics{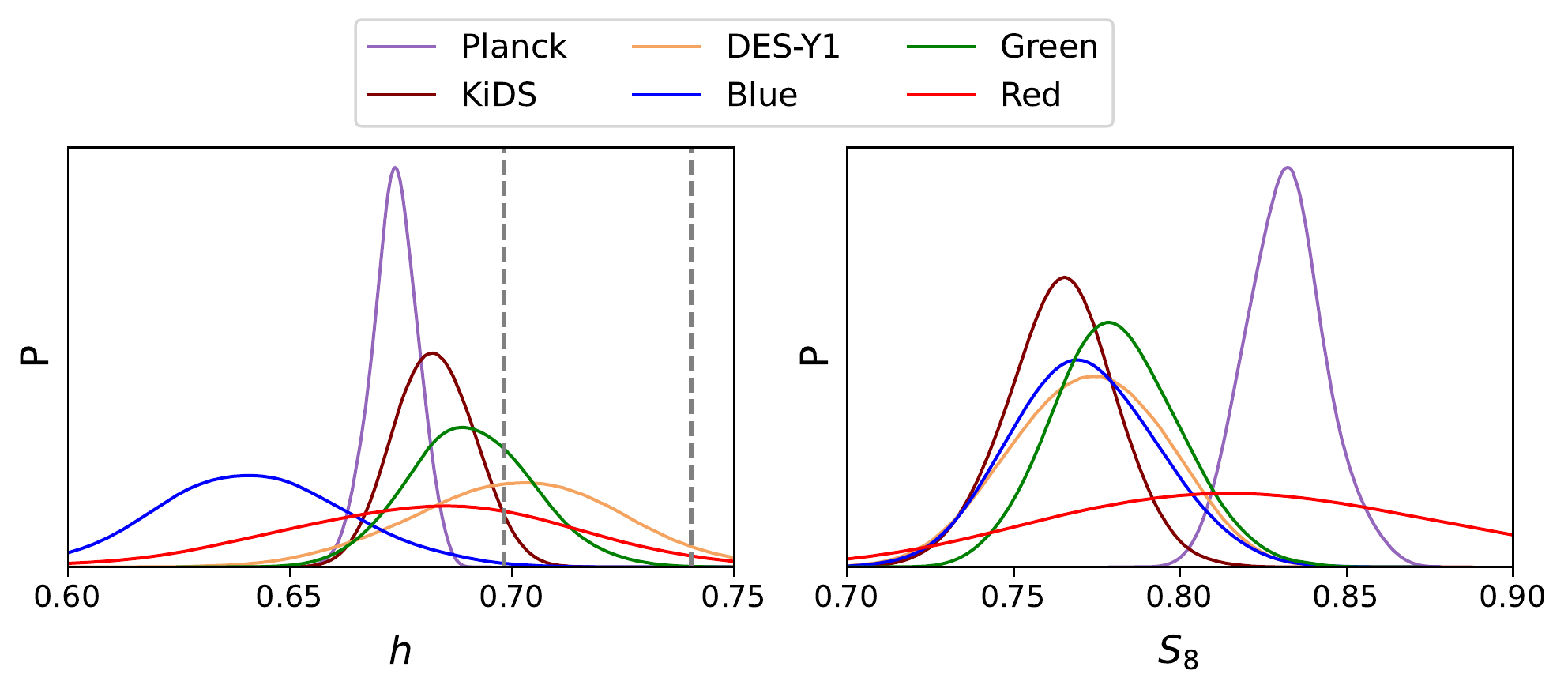}}
    \caption{Marginalized posteriors for the dimensionless Hubble constant, $h$, and lensing amplitude, $S_8=\sigma_8(\Omega_m/0.3)^{0.5}$.  Curves are shown for the three unWISE samples (blue, green and red), the Planck data, the KiDS+BOSS+2dFlens data and the DES Y1 data.  The DES and KiDS data alone do not meaningfully constrain $h$, but we have imposed a prior on $\Omega_m h^3$ as we did for the unWISE data and this leads to a constraint on $h$ via the DES or KiDS constraint on $\Omega_m$.  The vertical dashed lines in the left panel show the central values measured by the Carnegie-Chicago Hubble Program (left; \cite{Freedman19}) and the SH0ES team (right; \cite{Riess19}).
    }
    \label{fig:data_1d_posterior}
\end{figure}

\begin{figure}
    \centering
    \resizebox{\columnwidth}{!}{\includegraphics{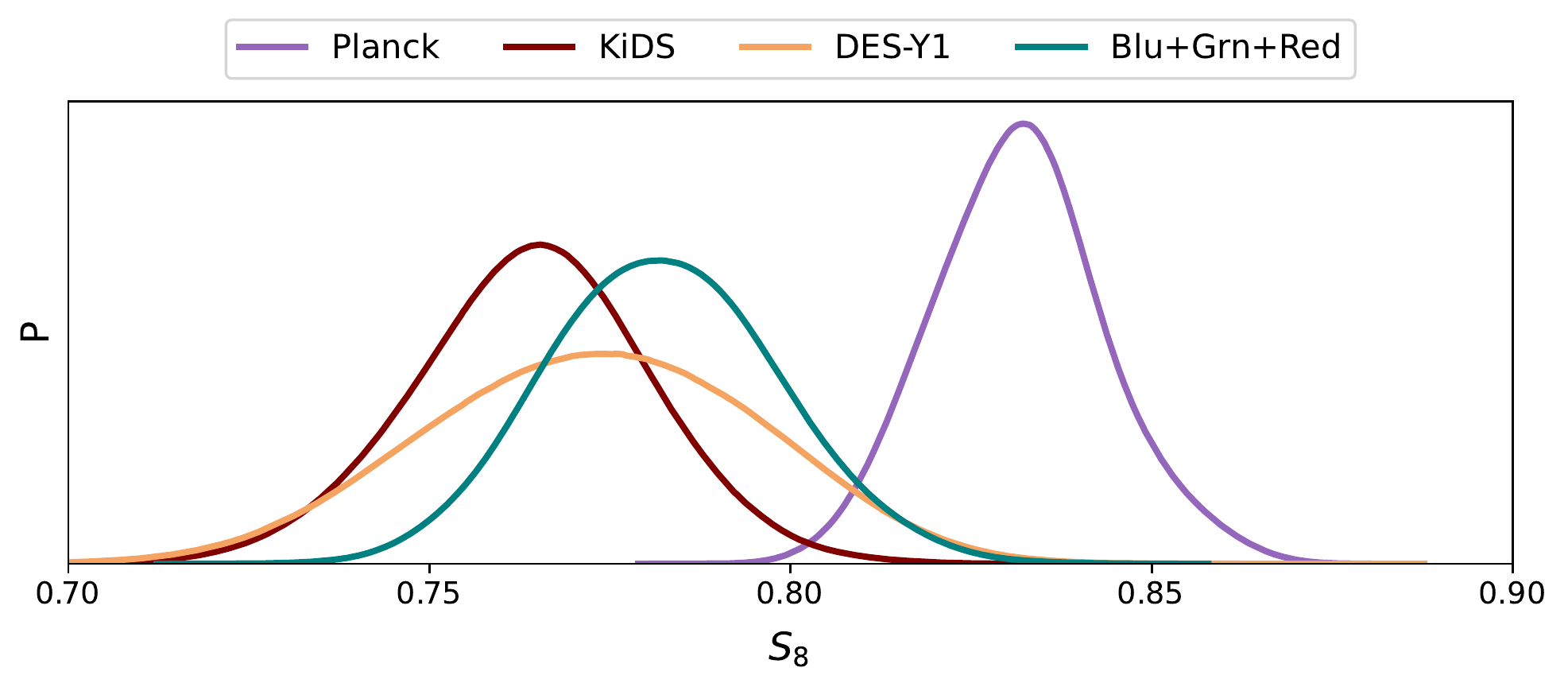}}
    \caption{Marginalized posteriors for the lensing amplitude, $S_8=\sigma_8(\Omega_m/0.3)^{0.5}$, from a variety of experiments.  Curves are shown for the Planck (purple \cite{PlanckParams18}), KiDS+BOSS+2dFlens (maroon \cite{kids1000}) and DES Y1 (light brown \cite{Abbott:2017wau}) data plus the combination of the blue, green and red unWISE samples (black; see text).
    }
    \label{fig:data_cmp_S8}
\end{figure}

The underlying cause of the $S_8$ tension is thus unclear.  Fortunately, both CMB and low-redshift measurements will rapidly improve over the next few years, which will either resolve the tension or more definitely isolate its source:
CMB data from ACT and SPT will soon have comparable statistical power to Planck, providing a great independent check, while Simons Observatory primary CMB measurements will halve the errors on $\Omega_m$ and $h$. Upcoming CMB lensing maps will allow sub-percent measurements of cross-correlations, which together with cosmic shear and clustering measurements from DES, DESI and Rubin Observatory (as well as upcoming space missions such as the Roman Space Telescope and Euclid) will shed light on the composition and expansion history of our Universe.

In Fig. \ref{fig:data_1d_posterior} we show our posterior on $h$, derived from the constraint that we keep $\Omega_m h^3$ fixed in the chains. For our fiducial blue+green+red sample we find $h = 0.68\pm 0.01$, slightly larger than Planck's $h = 0.677\pm 0.004$ (when including CMB lensing and low-redshift BAO).  
Adding low-redshift BAO to our fiducial blue+green+red sample
tightens the constraint on $h$ and shifts it towards Planck, $h = 0.68 \pm 0.005$.
 The value from the red is very similar, with approximately double the error.  Our fiducial result is fully consistent with the direct supernovae measurement by the Carnegie-Chicago Hubble Program \cite{Freedman19}, but still in mild tension with the SH0ES results \cite{Riess19} (shown as dashed lines in Fig.~\ref{fig:data_1d_posterior}).

\section{Conclusions}
\label{sec:conclusions}

The combination of low redshift galaxy surveys with CMB lensing provides a particularly powerful tool for constraining cosmology.  In an earlier paper \cite{Krolewski20} we presented the galaxy auto and galaxy-convergence cross spectra for three galaxy samples (at $z\simeq 0.6$, $1.1$ and $1.4$) selected from the unWISE catalog.  In this paper we use these spectra to constrain cosmological models, with an analysis that is fully blinded to the cosmology.  Our analysis is complementary to, but probes the same quantities as, the cosmic shear analyses of BOSS, KiDS and DES \cite{Leauthaud:2016jdb, kids1000, Abbott:2017wau} and because it uses the CMB as a source it is immune to intrinsic alignments of galaxies, blending or photometric redshift errors of the source galaxies.  The high redshift of our source screen also allows us to measure the lensing signal significantly above $z\simeq 1$, which is very difficult to do with cosmic shear.

To fit the spectra we employ a hybrid model that combines a fitting function for the matter power spectrum with Lagrangian perturbation theory calculations of scale-dependent bias (Section \ref{sec:model}).  Comparison with mock catalogs (Section \ref{sec:mocks}) shows that our model is capable of fitting data similar to our samples up to $\ell_{\rm max}\simeq 250-300$ (Section \ref{sec:testmodel}) with negligible biases.  While we have previously published the angular power spectrum measurements (and so are ``unblinded'') we performed all of our analysis blind to the cosmological parameters and did not change any of our procedures after unblinding.

We find good constraints on the lensing parameter, $S_8=\sigma_8(\Omega_m/0.3)^{0.5}$, and weaker constraints on the matter density, $\Omega_m$, and amplitude of clustering, $\sigma_8$ (Section \ref{sec:cosmology}; Figs.~\ref{fig:data_corner_plot}, \ref{fig:data_sml_corner_plot} and Table \ref{tab:data}).  Since we impose a constraint on $\Omega_m h^3$ (Section \ref{subsec:params_marg}) from the highly robust, preferred angular scale in the CMB our constraint on $\Omega_m$ implies a constraint on $H_0$ (Fig.~\ref{fig:data_1d_posterior}).  Our parameter constraints are broadly consistent with the previous literature, but confirm the tendency of lower redshift lensing measurements to give a lower $S_8$.  Our fiducial constraint (Fig.~\ref{fig:data_cmp_S8}) comes from a combination of the blue and green samples, which implies an $S_8$ that is just under $2.5\,\sigma$ lower than the value preferred by Planck and in good agreement with recent cosmic shear \cite{kids1000,Abbott:2017wau} and galaxy-galaxy lensing \cite{Singh20} measurements (both also at low redshift).  This agreement, with a method that is independent of many of the possible sources of systematic errors in those measurements, provides a valuable cross-check on the reported tension. While we have not attempted to combine different experiments, the agreement of several statistically independent measurements on a low value of $S_8$ is a strong indication of either a common systematic\footnote{We have reduced the set of possible common systematics significantly by using CMB lensing rather than galaxy lensing. Nonetheless, some of the non-linear modeling and photo-z methods are quite similar between different analyses.}, a fluctuation in the CMB measurements, or new physics between the surface of last scattering and the low-redshift Universe.

While our measurements and analyses are currently state-of-the-art in terms of signal to noise ratio and precision, we anticipate rapid progress in this field in the very near future.  First, combining more sensitive, ground-based measurements of the CMB with the existing Planck data should improve the signal to noise on $C_\ell^{\kappa g}$ for all samples, and hence the constraints on the power spectrum amplitude, $\sigma_8$.  Lower noise observations will increase the constraining power of the high $\ell$ data (that are not very constraining in our measurement) and require improvements in modeling the signals.  For such analysis we could employ a Lagrangian bias emulator \cite{Modi20,Kokron21,Zennaro21,Hadzhiyska21} or the HZPT model \cite{Sullivan21} which should both be sufficient to model all scales to $k_{\rm max}\simeq 0.6\,h\,{\rm Mpc}^{-1}$ or $\ell_{\rm max}=900$ at $z=0.6$.  Finally, the uncertainty on the redshift distribution, which currently contributes a large fraction of our error budget, can be reduced using spectroscopic observations of a subsample of the unWISE galaxies and further work on incorporating the uncertainty in $dN/dz$ could pay dividends.

\section*{Acknowledgments}
We thank Gerrit Farren, Catherine Heymans, Emmanuel Schaan, Eddie Schlafly, David Schlegel, Uro\v{s} Seljak, Blake Sherwin, An\v{z}e Slosar and David Spergel for useful discussion.
A.K.~is supported by the AMTD Foundation.
S.F.~is supported by the Physics Division of Lawrence Berkeley National Laboratory.
M.W.~is supported by the U.S.~Department of Energy and by NSF grant number 1713791.
We acknowledge the use of \texttt{NaMaster} \cite{Alonso18}, \texttt{Cobaya} \cite{CobayaSoftware, Cobaya}, \texttt{GetDist} \cite{Lewis:2019xzd}, \texttt{CAMB} \cite{Lewis:1999bs} and \texttt{velocileptors} \cite{Chen20} and thank their authors for making these products public. This research used resources of the National Energy Research Scientific Computing Center (NERSC), a U.S.\ Department of Energy Office of Science User Facility operated under Contract No.\ DE-AC02-05CH11231.
This research was enabled in part by software provided by Compute Ontario (\url{https://www.computeontario.ca}) and Compute Canada (\url{http://www.computecanada.ca}).
This work made extensive use of the NASA Astrophysics Data System and of the {\tt astro-ph} preprint archive at {\tt arXiv.org}.

\appendix

\section{CrowCanyon2 simulations}
\label{app:crowcanyon2}

The CrowCanyon2 simulations were created using the FastPM code \cite{FengChuEtAl16}, modified to enhance performance on the NERSC\footnote{http://www.nersc.gov} machine Cori.
Each run evolved $8192^3$ equal mass particles in a periodic, cubic volume of $4096\,h^{-1}$Mpc resulting in a mean inter-particle separation of $0.5\,h^{-1}$kpc (comoving) and a mass resolution of $1.1\times 10^{10}\,h^{-1}M_\odot$.  The code used a $24576^3$ mesh  (i.e.~a force resolution factor $B=3$) for the gravity calculation and took 40 time steps from $z=19$ to $z=0$.   The linear power spectrum was computed using \texttt{CAMB} \cite{Lewis00} at $z=0$ and initial conditions were generated from this using second-order Lagrangian perturbation theory at $z=19$.

\begin{figure}
    \centering
    \resizebox{\textwidth}{!}{\includegraphics{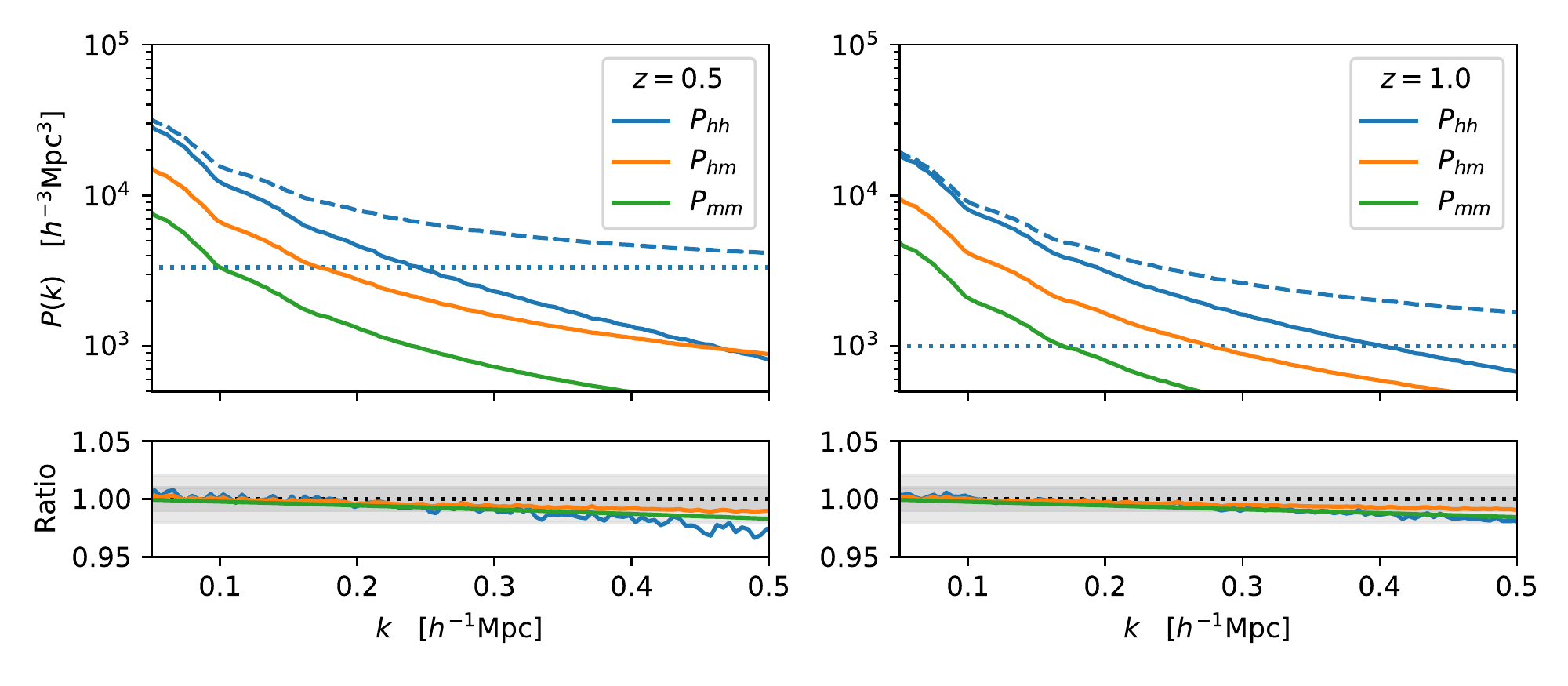}}
    \caption{Comparison of the real-space power spectra between the TreePM and FastPM simulations for two example redshifts ($z = 0.5$ and $z=1.0$) and number densities ($\bar{n}=3\times 10^{-4}\,h^3\mathrm{Mpc}^{-3}$ and $\bar{n}=10^{-3}\,h^3\mathrm{Mpc}^{-3}$).  The upper panels show the halo and matter auto- and cross-power spectra, while the lower panels show the ratio FastPM/TreePM for each statistic (the light and dark grey bands showing 1 and 2\% agreement).  For the halo autospectrum in the upper panels we show the results with (solid) and without (dashed) Poisson shot-noise subtracted (shown as the horizontal, dotted line in the upper panels).  In the lower panels the ratio is with shot-noise subtracted.}
    \label{fig:cmp_pkr}
\end{figure}

To demonstrate the accuracy of the chosen configuration we compare FastPM simulations to four TreePM \cite{TreePM} simulations.  The TreePM code has been compared to other N-body codes in ref.~\cite{Heitmann08}, finding percent level agreement on basic statistics.  We used existing TreePM runs with similar mean interparticle spacing and volume sufficient to probe the lengths and halo masses of interest \cite{Reid14,White15b}.  Specifically we started four FastPM runs from the initial conditions of four TreePM simulations, each employing $2048^3$ particles in a box of $1380\,h^{-1}$Mpc.  These simulations differed only in the random number seed chosen for the initial conditions and we used the average of the four comparisons to reduce the statistical noise.  As for the CrowCanyon2 runs, we used $B=3$ and 40 time steps from $z=9$ to $z=0$.  Due to the ratio of box size to particle load this is slightly lower resolution than the CrowCanyon2 configuration, so the convergence test should be considered conservative.
Across a wide range of number densities and redshifts appropriate
for the unWISE samples, we find agreement to better than 1\% between the halo
autospectrum and halo-matter cross-spectrum
of the two simulation boxes to $k = 0.3$ $h$ Mpc$^{-1}$,
the scales of interest for the unWISE analysis (Fig.~\ref{fig:cmp_pkr}).

We measure $C_{\ell}^{gg}$ and $C_{\ell}^{\kappa g}$
on the lightcone to create realistic mocks.
A particle at space-time coordinate $(\mathbf{x}(t), t)$ is included in the light-cone, if and only if the distance from the particle to the observer is the same as the distance light can travel for the same duration, also known as the co-moving distance $D_c$,
\begin{equation}
  D_c(t_\mathrm{emit}) = | \mathbf{x}_i(t_\mathrm{emit})|.
\label{eq:lightcone}
\end{equation}
In the simulation, the particle trajectory $x_i(t)$ is interpolated between time steps using the FastPM Kick and Drift factors.
The simulation box is tiled to allow full-sky coverage
of the lightcone.
On average, each light-cone slice contains the same number of particles as the total number of particles used in the simulation.

Convergence maps were generated as a weighted
surface mass density
\cite{BartelmanSchneider01}
\begin{equation}
    \kappa(\theta) = \frac{3 H_0^2 \Omega_m}{2} \int_0^{z_*} \frac{dz}{H(z)} \frac{(1+z) \chi(z) (\chi_{\star} - \chi(z))}{\chi_*} \delta_m (\chi \theta, \chi; z)
\end{equation}
where $\chi_\star$ is the comoving distance
to the surface of last scattering.  This implementation
is available as part of the \textsc{SIMPLEHOD} package\footnote{\url{https://github.com/bccp/simplehod/blob/master/scripts/wlen.py}}, which also includes
the implementation of the unWISE HOD.
We also generate magnification maps for each of the three samples,
$\mu_i(\theta)$, in the same way:
\begin{equation}
    \mu_i(\theta) = \frac{3 H_0^2 \Omega_m}{2} (5s_{\mu_i} - 2) \int_0^{z_*} \frac{dz}{H(z)} \int_z^{z_*} \frac{dz'}{H(z')} \frac{\chi(z) (\chi(z') - \chi(z))}{\chi(z')} H(z') \frac{dN_i}{dz'}
    \delta_m (\chi \theta, \chi; z)
\end{equation}

\clearpage

\section{Posterior distributions in mocks and data}
\label{app:posteriors}

\begin{figure}
    \centering
    \includegraphics[width=0.49\textwidth]{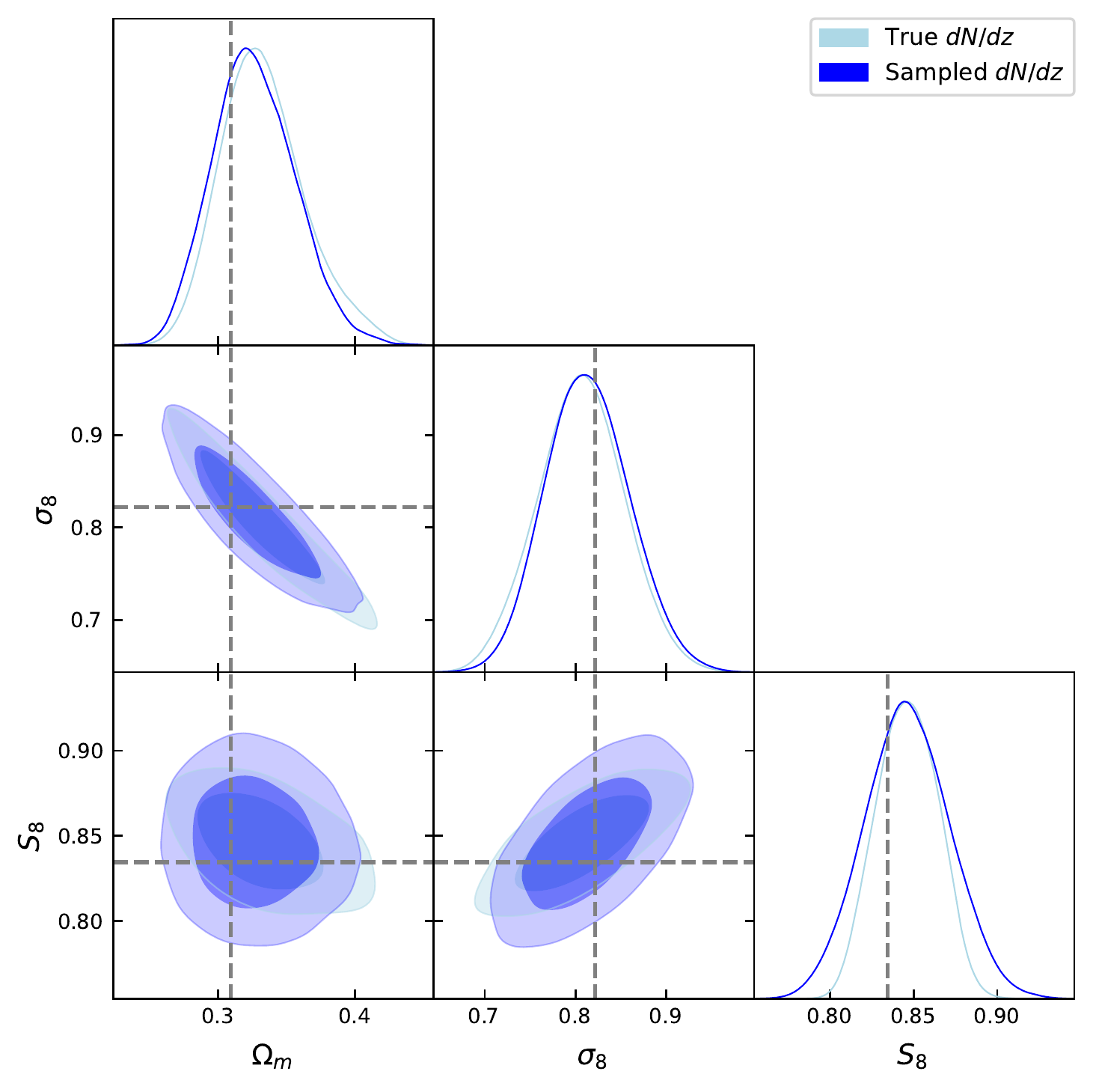}
    \includegraphics[width=0.49\textwidth]{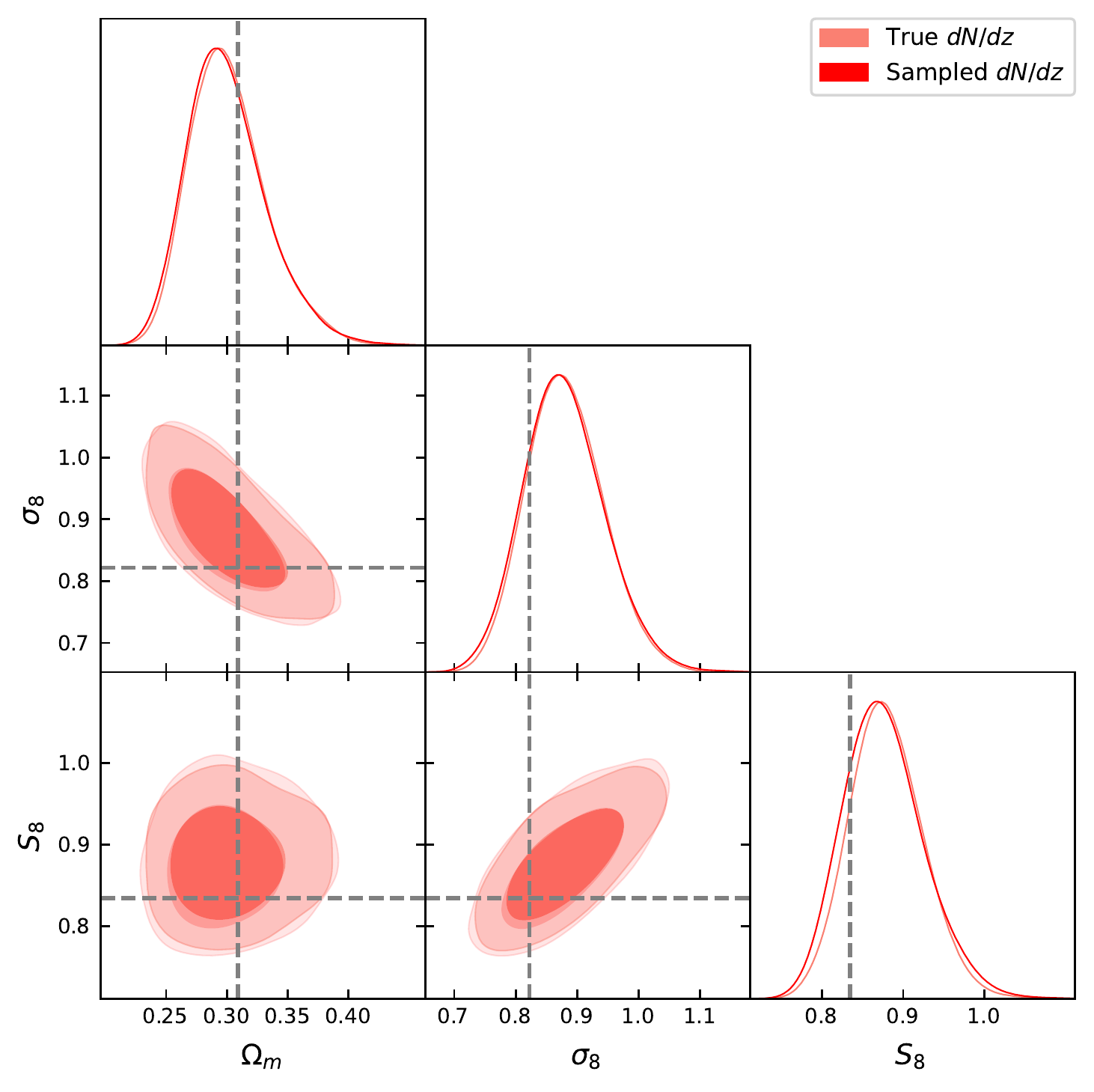}
    \caption{Posteriors for the mock blue (left) and red (right) samples, showing the key cosmological parameters: $\Omega_m$ and $\sigma_8$ plus the derived parameter $S_8$.  Results for both fixed $dN/dz$ and sampling over the $dN/dz$ uncertainty are shown (see discussion in text).  The dashed grey lines show the ``true'' values of the parameters in the simulation, though we expect some scatter due to sample variance in the initial conditions. }
    \label{fig:mock_two_contours}
\end{figure}

Fig. \ref{fig:mock_two_contours} shows the posteriors for the mock blue and red samples, which are also unbiased under our modeling assumptions.

\begin{figure}
    \centering
    \includegraphics[width=\textwidth]{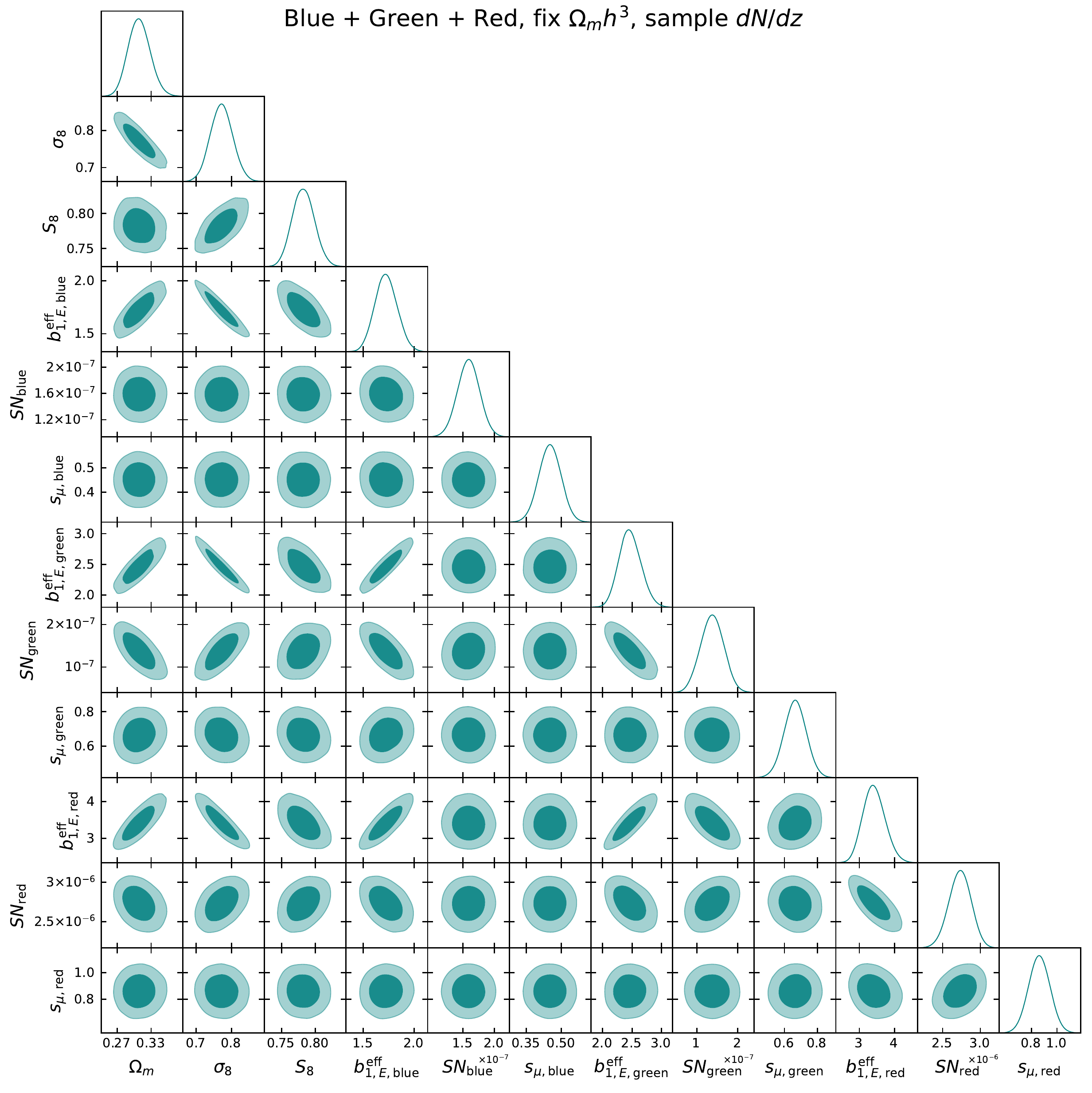}
    \caption{Posteriors for the full set of cosmological and nuisance parameters for the blue + green sample.  The meanings of each parameter are discussed in the main text. }
    \label{fig:big_corner_plot}
\end{figure}

Fig. \ref{fig:big_corner_plot} shows the posteriors for the full set of cosmological and nuisance parameters for our fiducial, blue + green, sample.

\clearpage

\bibliographystyle{JHEP}
\bibliography{main}

\end{document}